\documentclass[12pt]{article}
\usepackage[english]{babel}
\usepackage[pdftex]{graphicx}
\usepackage[authoryear,round]{natbib}
\usepackage[margin=20mm]{geometry}
\usepackage{hyperref}
\hypersetup{
    colorlinks = true,
    linkcolor = blue,
    anchorcolor = blue,
    citecolor = blue,
    filecolor = blue,
    urlcolor = blue,
	pdfauthor={some author},
	pdftitle={eye-catching title}
    }
\usepackage{amssymb,amsmath}
\usepackage{url}

\begin{document}

\title{The growing family of gamma-ray\\narrow-line Seyfert 1 galaxies\footnote{Presented at: \emph{Observations and physics of NLS1 galaxies: AGN at their extreme}, November 26-28, 2025, European Southern Observatory, Vitacura, Santiago (Chile).}}
\author{Luigi Foschini\footnote{Brera Astronomical Observatory, National Institute of Astrophysics (INAF), Milano/Merate, Italy}.}
\date{January 26, 2026}
\maketitle

\begin{abstract}
The revision of the fourth \emph{Fermi} Large Area Telescope (LAT) catalog of gamma-ray point sources (rev4FGL) revealed that the gamma-ray sky is populated by emerging populations of jetted active galactic nuclei (AGN) other than blazars and radio galaxies. Narrow-Line Seyfert 1, Seyfert 1, intermediate, and Seyfert 2 galaxies, changing-look AGN, plus a number of ambiguous or unclassified sources. After a short historical introduction on the gamma-ray observations of Seyfert-type AGN, I explore the main statistical properties of 1477 jetted AGN from the rev4FGL with spectroscopic redshift, and also the cross-match with Very Large Baseline Array (VLBA) radio observations at 15~GHz from the Monitoring Of Jets in Active galactic nuclei with VLBA Experiments (MOJAVE) program. I then discuss the difference between gamma and non-gamma jetted AGN, and the implications on the classification. 
\end{abstract}

% Nota: rammentare di usare sempre radio galaxy e non radiogalaxy!

\section{Introduction}
The superb performance of the Large Area Telescope (LAT) onboard the \emph{Fermi Gamma-Ray Space Telescope} ($2008-$today, \citealt{ATWOOD2009}) made it possible to observe the gamma-ray sky with unprecedented sensitivity. The latest version of the fourth \emph{Fermi}/LAT catalog (4FGL-DR4, \citealt{BALLET2023}) contains 7194 gamma-ray point sources collected in fourteen years of observations. To grasp the jump in performance, one can compare the 4FGL with the 271 gamma-ray point sources of the third, and last, catalog of the Energetic Gamma Ray Experiment Telescope (EGRET) onboard the \emph{Compton Gamma-Ray Observatory} \citep{HARTMAN1999}.  

Among the novelties introduced by \emph{Fermi}/LAT, I would like to focus here on the Seyfert-type active galactic nuclei (AGN). Already during the early months of operation, it was possible to discover high-energy gamma rays from the relativistic jet of the narrow-line Seyfert 1 galaxy (NLS1) PMN~J$0948+0022$ ($z=0.585$, \citealt{ABDO2009B,ABDO2009D,FOSCHINI2010}), and after one year of operation, the number of gamma-ray NLS1s increased to four (\citealt{ABDO2009E}; see also \citealt{FOSCHINI2012,FOSCHINI2020} for reviews on this topic). After the revision of the second data release (DR2) of the 4FGL \citep{FOSCHINI2021,FOSCHINI2022}, the number of gamma-ray NLS1s is today around two dozens, but the revision also showed that other Seyfert-type AGN are gamma-ray emitters. I would like to emphasize that this is not a mere question of classification, of sticking a label on a cosmic object, but a question that hides deep physical problems. As I have already underlined (e.g. \citealt{FOSCHINI2017}), the key point is the relatively small mass of the central black hole, the high accretion rate and the disk host galaxy, because this implies a breakdown of the paradigm of blazars and radio galaxies (e.g. \citealt{BLANDFORD1990}), and can therefore help us understand the formation and physics of relativistic jets. 

Broadening the discussion to include the entire Seyfert class also implies extending research further back in time. While NLS1s were formally defined in 1985 \citep{OSTERBROCK1985}, the first seminal paper on the Seyfert galaxies dates back to 1943 \citep{SEYFERT1943}. This is particularly important with respect to the detection and properties of gamma rays, since, as I will show later, the search for gamma rays emitted by Seyferts is older than that by NLS1.

There is one more reason to consider the entire Seyfert class. Indeed, there is an increasing awareness of time evolution -- on human timescales -- of AGN, which affects significantly their observational appearance, and therefore their classification. This phenomenon is generally called changing-look AGN (for a recent review see \citealt{RICCI2023}; for jetted AGN see \citealt{FOSCHINI2021,FOSCHINI2022,RAITERI2025}) and refers to changes in the observational appearance of the same AGN: changes in the emission lines (presence/absence of broad-emission lines), changes in the obscuration (Compton thick/thin), changes due to the activity of the relativistic jet (line-dominated/featureless optical spectra). Here are a couple of examples, crucial for this research. \cite{DALLABARBA2025} studied a series of optical spectra of the gamma-ray emitting NLS1 PMN~J$0948+0022$ showing an apparent change in the H$\beta$ emission line profile, from a Lorentzian shape (typical of NLS1s) to a composite Gaussian one (broad component plus a narrow peak, typical of intermediate Seyferts). However, this change was due to the interaction of the relativistic jet with the narrow-line region, and not to a misclassification. \cite{LAHTEENMAKI2018} reported about strong (Jy-level) radio outburst detected at 37~GHz from radio-quiet and -silent NLS1s, making it clear that the radio-quiet/-loud distinction is more useless than ever. All these discoveries raise obvious doubts about the boundaries between the different subclasses of the Seyfert family and of the AGN in general. 

Considering that the boundaries between the various Seyfert subclasses are now quite blurred, I would like to start these notes by expanding the topic to all the Seyfert galaxies, which in turn requires approaching the problem from the high-energy part of the electromagnetic spectrum. 

\begin{figure}[!t]
\begin{center}
\includegraphics[scale=0.15]{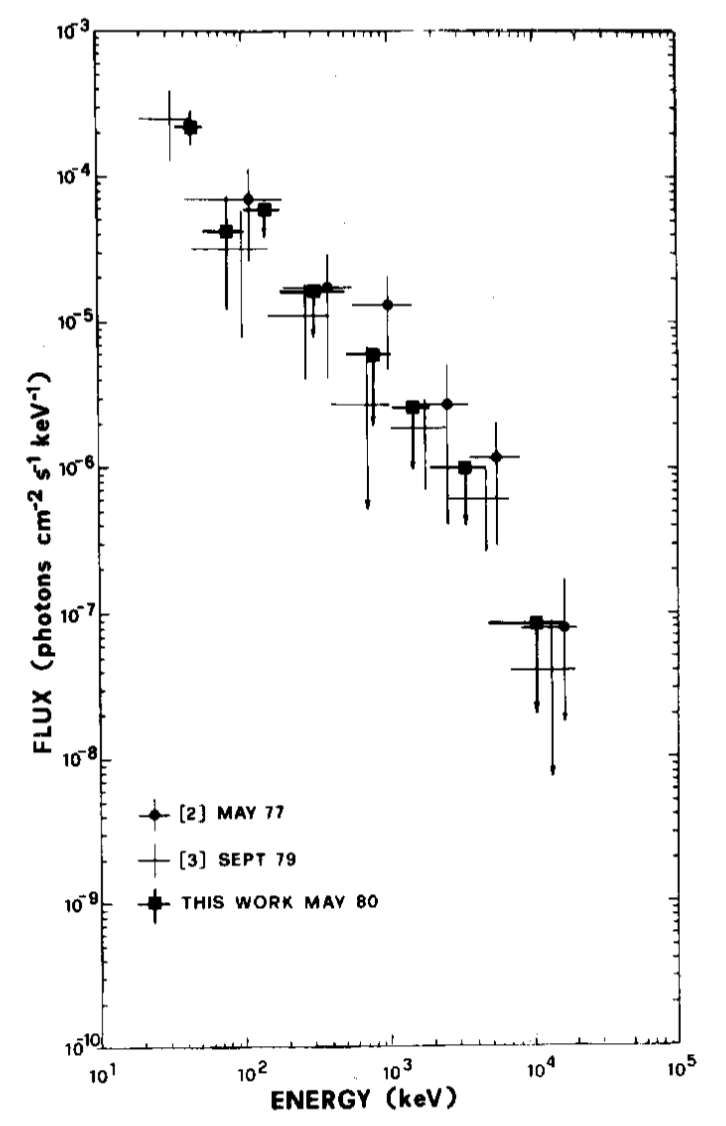}
\includegraphics[scale=0.15]{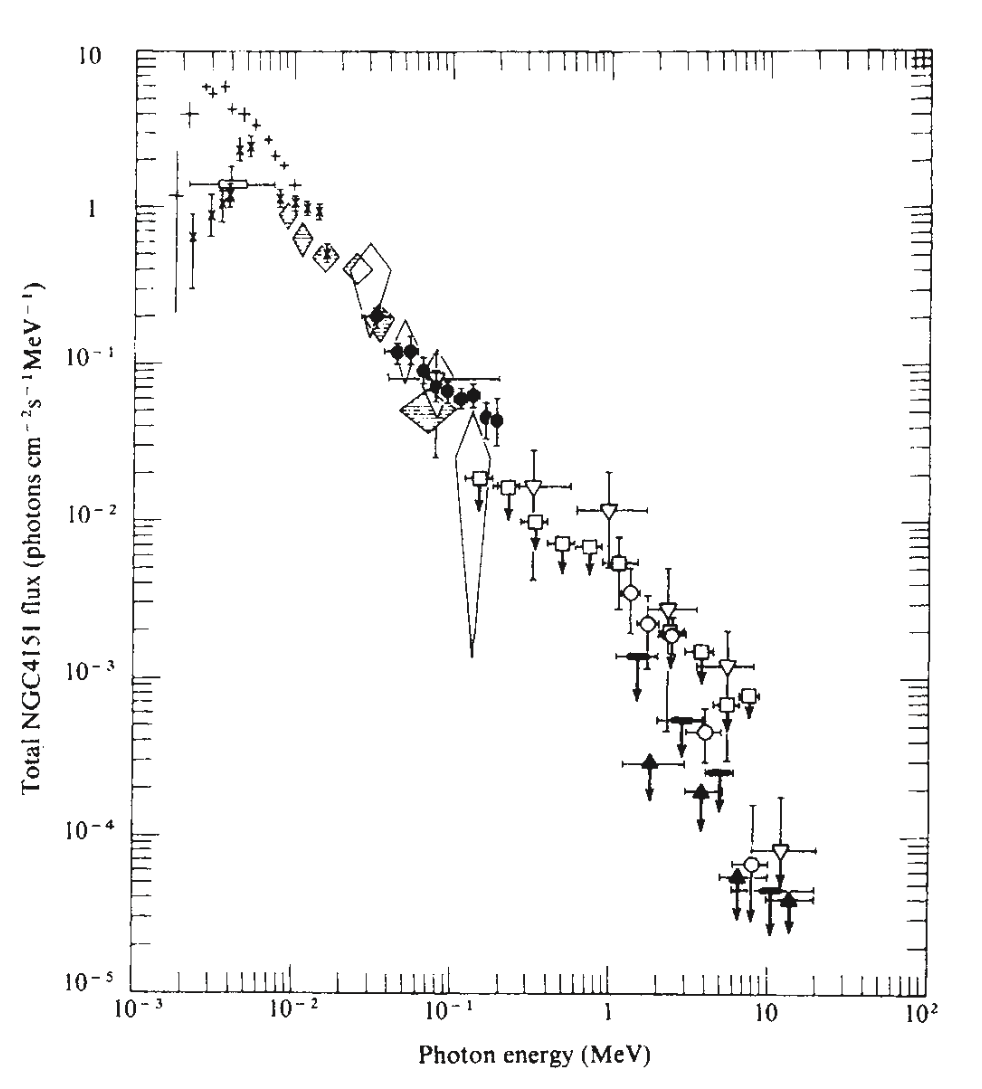}
\includegraphics[scale=0.15]{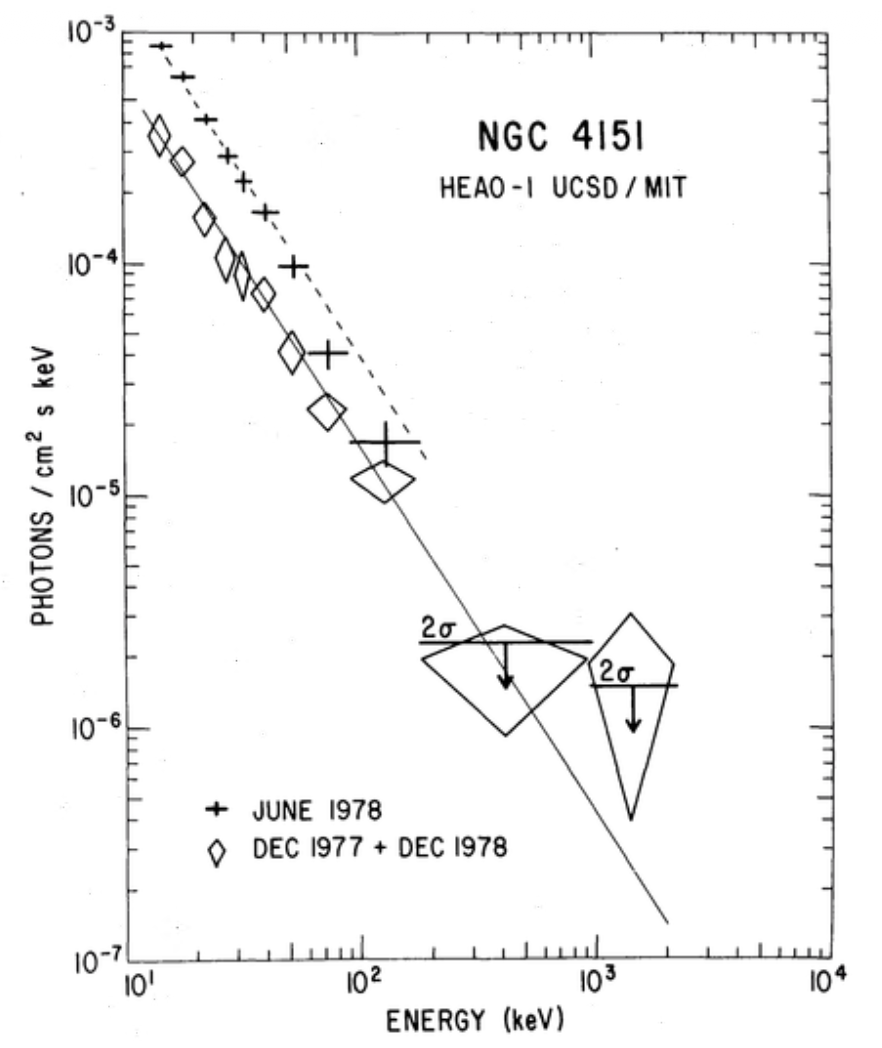}
\caption{Energy spectrum of the type-1.5 Seyfert NGC 4151 (\emph{left panel} from \citealt{PEROTTI1983}; \emph{center panel} from \citealt{WHITE1980}; \emph{right panel} from \citealt{BAITY1984}).}
\label{fig:ngc4151}
\end{center}
\end{figure}

\section{Early gamma-ray observations of Seyfert galaxies}
The United States \emph{Small Astronomy Satellite 2 (SAS2)}, carrying one spark chamber operating in the $0.02-1$~GeV energy band, was launched on November 19, 1972.  Although a failure of the power supply ended prematurely the mission on June 8, 1973, \emph{SAS2} proved the existence of a gamma-ray sky, with Galactic and extragalactic components \citep{FICHTEL1975}. 

The search for gamma rays from Seyferts started already during 70s: it was expected that photons up to $\sim 1$~GeV could avoid pair production in this class of AGN \citep{BERGERON1971,HERTERICH1974}. In 1977, the balloon-borne Milan-Southampton (MISO) gamma-ray telescope ($0.2-20$~MeV) detected the Seyfert type-1.5 NGC~4151 ($z=0.00315$, \citealt{DICOCCO1977,PEROTTI1979}). This result was initially confirmed by the balloon-borne Max-Planck-Institut (MPI) Compton telescope ($1-20$~MeV, \citealt{GRAML1978}), but this work was later withdrawn\footnote{Personal communication by V. Sch\"onfelder to R. White \citep{WHITE1980}, reported also by \cite{DEAN1981}.}. However, the detection was confirmed later by other MISO observations \citep{PEROTTI1981A,PEROTTI1983}, and by the A4 Medium Energy Detectors (MEDs, $0.025-2.1$~MeV) onboard the \emph{High Energy Astronomy Observatory 1} (HEAO 1, \citealt{BAITY1984}). In these latter cases, there were also non-detections, suggesting a strong variability of the gamma-ray emission (Fig.~\ref{fig:ngc4151}, see also \citealt{GEHRELS1992}). These results were challenged on basis of the wide field of views of the instruments (some degrees, implying the possibility of contaminating sources), and the uncertainties on the diffuse gamma-ray background (e.g. \citealt{WHITE1980,GRASER1982,ZDZIARSKI1996}). \cite{PEROTTI1979} found a positional uncertainty of $0.8^{\circ}$ in right ascension, with most of emission within $\sim 20'$ of NGC~4151 and reported only one nearby source, the BL Lac Object 2A~$1219+305$ at $\sim 9.5^{\circ}$ away (see also \citealt{DEAN1981}). Later, they reported another closer X-ray source, at $\sim 5'$ (\citealt{PEROTTI1981A}; see also the discussion in Sect.~\ref{fermisky}). There was also a claim of detection above 100~MeV by \cite{GALPER1979}: they used a spark chamber ($0.1-1.5$~GeV, field of view $\sim 10^{\circ}$) onboard the \emph{Kosmos 264} satellite that scanned the area around NGC~4151 in 1969 and measured a flux excess of $4.4_{-1.7}^{+2.1}\times 10^{-5}$~ph~cm$^{-2}$~s$^{-1}$, which is a quite huge value for an extragalactic object, at level of the Vela pulsar.

In 1978, the European \emph{Celestial Observation Satellite B (COS-B)} detected high-energy gamma rays ($50-500$~MeV) from the flat-spectrum radio quasar (FSRQ) 3C~273 ($z=0.157$, \citealt{SWANENBURG1978}). Later, \cite{STRONG1983} suggested that NGC~1275 (a.k.a. Per A) can be the source of a gamma-ray excess observed by \emph{COS-B} while pointing at the Perseus cluster. 

Balloon-borne experiments continued, leading to new detections of another Seyfert, MCG~$8-11-11$ ($z=0.0201$, \citealt{PEROTTI1981B}, up to $\sim 3$~MeV), and the nearby radio galaxy Centaurus A ($z=0.00188$, \citealt{BAITY1981,VONBALLMOOS1987}, up to $\sim 20$~MeV). There was also a claim of MeV detection of the Seyfert 1 3C~120 ($z=0.0336$) by \cite{DAMLE1986} by using the spark chamber Natalya-1 ($5-100$~MeV) onboard a ballon. They reported a flux of $(3.6\pm1.2)\times 10^{-4}$~ph~cm$^{-2}$~s$^{-1}$, quite huge for an extragalactic source.

Two reviews of the time on extragalactic gamma-ray astronomy \citep{DEAN1981,BASSANI1983} classified Cen A already as a radio galaxy, while NGC~1275 was classified as unusual Seyfert galaxy, because it is hosted in a giant elliptical galaxy, and some authors even suggested that it had a BL Lac core. The spectral breaks observed at a few MeV in NGC~4151 and MCG~$8-11-11$, coupled with the upper limits from \emph{SAS2}, suggested that it might be a common characteristics of Seyferts and other emission-line active galaxies \citep{BIGNAMI1979}. The explored theories to generate gamma rays where based on the Penrose process for rotating black holes, synchrotron and Compton radiation, accretion onto black holes, and relativistic jets (e.g. \citealt{BERGERON1971,MARASCHI1979,PINKAU1980,MARSCHER1980,KONIGL1981,DEAN1981,BASSANI1983}). It is intriguing to note that \cite{BASSANI1983} proposed that part of the soft gamma-ray emission from  NGC~4151 might be due to a radio jet, which was discovered just in those years \citep{ULVESTAD1981,BOOLER1982,JOHNSTON1982}.

\begin{figure}[!t]
\begin{center}
\includegraphics[scale=0.3]{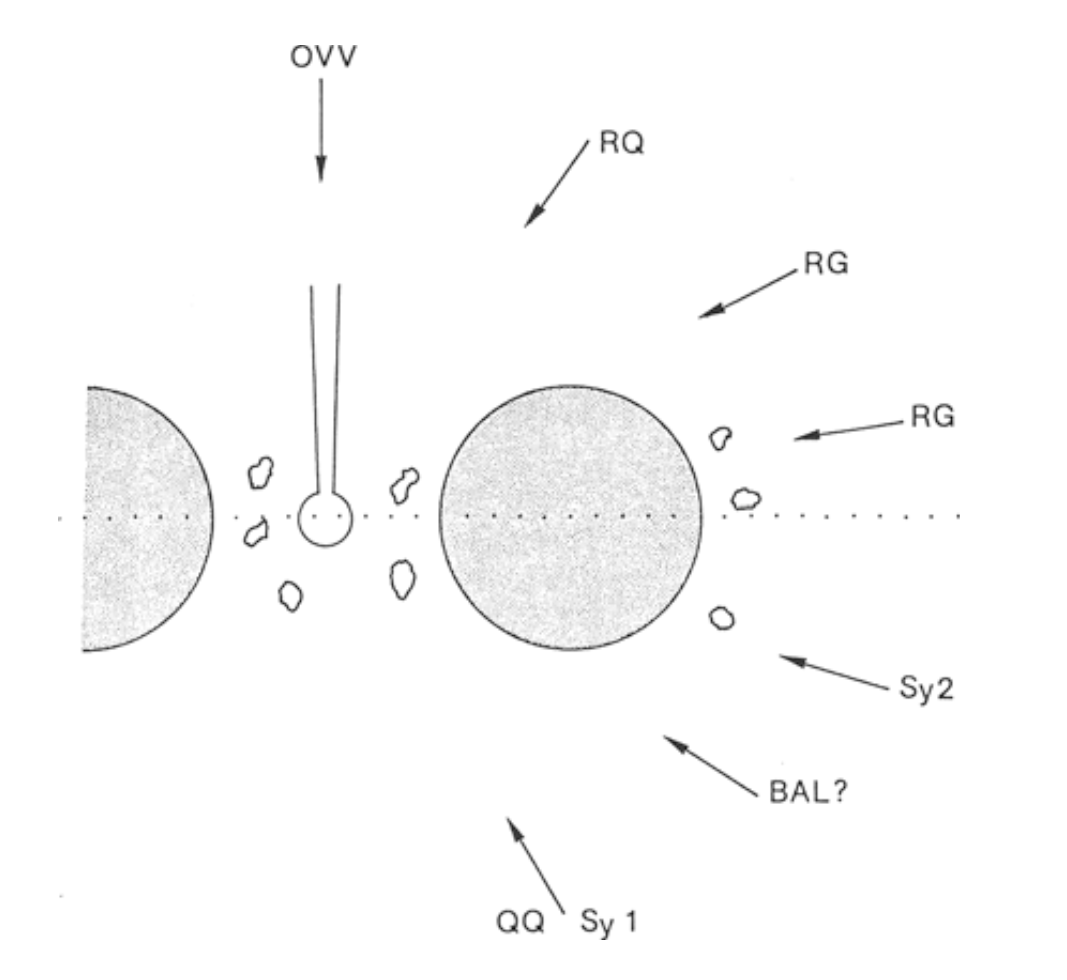}
\caption{The unified model of AGN outlined by L. Woltjer (from \citealt{BLANDFORD1990}). From the top to the bottom: Optically Violent Variable (OVV), Radio Quasars (RQ), Radio Galaxies (RG), Seyfert 2 galaxies (Sy2), Broad Absorption Lines quasars (BAL),  Seyfert 1 galaxies (Sy1), Radio Quiet Quasars (QQ). Interestingly, Woltjer had already noted the phenomenon of the changing-look of AGN in the text accompanying the figure, and cited the cases of NGC~$3516$, Fairall~$9$, PKS~$0521-36$.}
\label{fig:unifmod}
\end{center}
\end{figure}

The search for gamma-ray detection of Seyfert galaxies continued with the launch of the \emph{Compton Gamma-Ray Observatory (CGRO)} ($1991-2000$), which carried the Energetic Gamma Ray Experiment Telescope (EGRET) working in the $0.02-30$~GeV energy range. A sample of 22 Seyferts was selected on the basis of their X-ray flux, but no detection was found after about two years of observations \citep{LIN1993}. Of course, NGC~4151 was the main target, because of the MeV detections \citep{DICOCCO1977,PEROTTI1983,BAITY1984}, although later observations at hard X-rays with \emph{GRANAT}/SIGMA \citep{JOURDAIN1992} and \emph{CGRO}/OSSE \citep{MAISACK1993} have shown a rather steep spectrum with a break already at $\sim 100$~keV. \cite{JOURDAIN1992} and \cite{MAISACK1993} debated about a possible strong variability of NGC~4151 to explain the inconsistency between balloon-borne detections in the MeV energy range and the upper limits from hard X-rays (see also \citealt{PINKAU1980}). On the basis of the lack of \emph{CGRO}/EGRET detections (with upper limits of $\sim 10^{-7}$~ph~cm$^{-2}$~s$^{-1}$, about one order of magnitude smaller than those from \emph{SAS2} and \emph{COS-B}), \cite{LIN1993} concluded that the spectral break at hundreds of keV could be a common feature of Seyferts. Since \emph{CGRO}/EGRET already detected other jetted AGN (e.g. \citealt{DERMER1992}), \cite{LIN1993} also suggested that the lack of gamma rays from Seyferts can be due to the absence of jets, or, if present, viewed at large angles, according to the unified model of the time (Fig.~\ref{fig:unifmod}, \citealt{BLANDFORD1990}). 

However, it is very interesting to read the Lin's list of sources and understand what was meant by Seyfert galaxy. Most sources were type-1 Seyfert, but there were also two NLS1s (Mrk~335, NGC~4051). In particular, four sources deserve attention, because they were known to have relativistic jets from radio observations: the already cited NGC~1275, 3C~111, 3C~120, and 3C~390.3\footnote{3C~390.3 is the only jetted AGN in the list that has never been detected in gamma rays so far.}. According to the paradigm in the reference book by \cite{BLANDFORD1990} cited by Lin (Fig.~\ref{fig:unifmod}), the Seyferts were defined as AGN without jets hosted by spiral galaxies, although they can have some radio emission. The maximum limit for Seyferts of the radio emission was set by Woltjer at $\nu L_{\nu}(5\,{\rm GHz})=10^{41.4}$~erg~s$^{-1}$ on the basis of the selected sample of AGN. In his contribution, Blandford suggested that the different host galaxy can be the indication of a mass requirement to generate relativistic jets: ellipticals hosted high-mass black holes, while the opposite is for spirals.  

Back to the radio Seyferts of the Lin's list, NGC~1275 and 3C 111 are hosted in elliptical galaxies, and therefore should not be considered Seyferts according to the paradigm of the time, while the host of 3C~120 and 3C~390.3 are disk galaxies. Although the two latter satisfy the host requirement for a Seyfert, they easily exceed the radio power threshold defined by Woltjer. Therefore, for one reason or another, all these radio sources should not have been in Lin's list. However, \cite{GASKELL1987} had a different point of view saying that there is nothing that prevent a Seyfert to be a radio galaxy, and cited the case of NGC~1275. If it seems like there's a bit of confusion, the answer is yes! As noted by Roger Blandford in his contribution \citep{BLANDFORD1990}, ``the taxonomy of AGN is a very confused (and confusing) subject'', echoed by Martin Gaskell: ``It is obvious that these classes are very distance and instrument dependent'' \citep{GASKELL1987}. 

After the end of \emph{CGRO}, \cite{CILLIS2004} used all the available EGRET data and prepared two lists of objects by dividing radio galaxies from Seyferts. No detection was found, with the exception of the nearby radio galaxy Cen~A, which was already present in the third EGRET catalog \citep{HARTMAN1999}. However, once again, some radio galaxy has snuck into the Seyferts list, such as NGC~315 ($z=0.0167$) and NGC~6251 ($z=0.0230$), both giant radio galaxies with Mpc-scale jets. \cite{CILLIS2004} noted that NGC~6251 was proposed to be the counterpart of the gamma-ray source 3EG~J$1621+8203$ by \cite{MUKHERJEE2002}, but dismissed this association, believing it was a Seyfert with low radio brightness. It is also worth noting that \cite{MUKHERJEE2002} proposed the association on the basis of a partial coverage of the EGRET probability contour with \emph{ROSAT} and \emph{ASCA} archival X-ray observations. The complete coverage of the EGRET probability contours was done later by \cite{FOSCHINI2005}, who used the wide field of view of the IBIS imager onboard \emph{INTEGRAL} and confirmed the association of 3EG~J$1621+8203$ with NGC~6251.

Yet another Seyfert radio galaxy -- 3C~111 -- was associated to an EGRET source after the reanalysis of the data. \cite{HARTMAN2008} proposed that the gamma-ray source 3EG~J$0416+3650$ is composed by the contribution of three sources, one of them being 3C~111, which accounts for the gamma-ray emission above $\sim 1$~GeV. 

\begin{figure}[!t]
\begin{center}
\fbox{\includegraphics[scale=0.3]{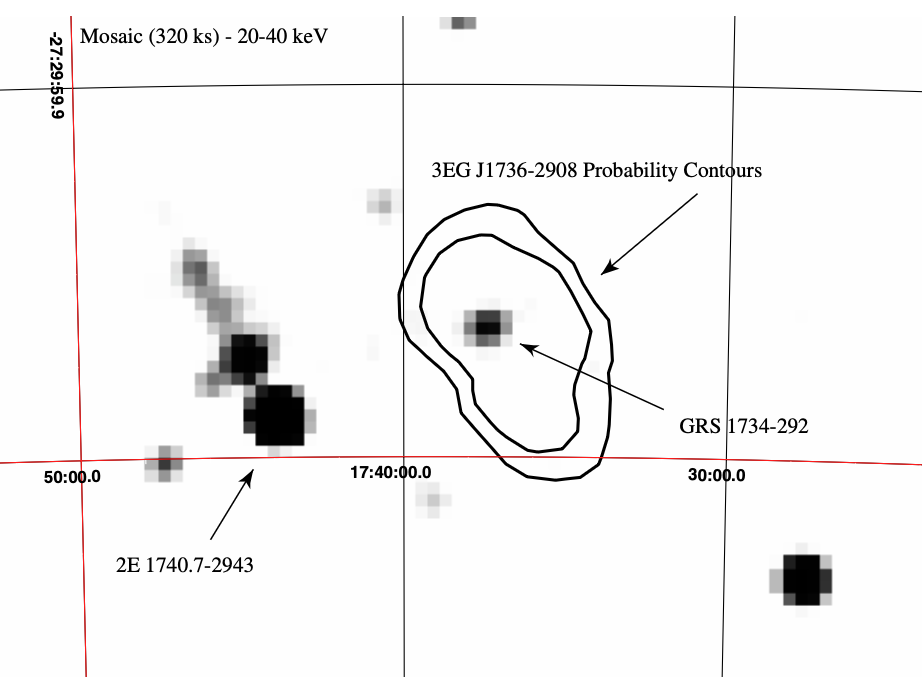}}
\includegraphics[scale=0.3]{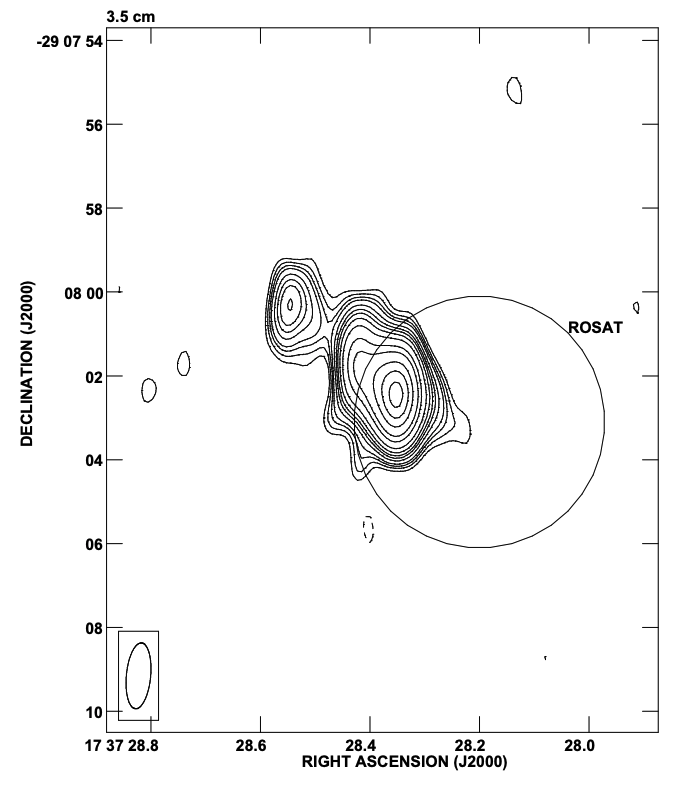}
\caption{(\emph{Left panel}) \emph{INTEGRAL}/IBIS image of the region around 3EG~J$1736-2908$ in the $20-40$~keV energy band, showing the 95\% and 99\% probability contours of the gamma-ray source superimposed (from \citealt{DICOCCO2004}). The Seyfert 1 galaxy GRS~$1734-292$ is the only source detected inside the contours. (\emph{Right panel}) Very Large Array (VLA) observation at $3.5$~cm ($8.5$~GHz) of GRS~$1734-292$, showing the jet-like structure (from \citealt{MARTI1998}).}
\label{fig:grs1734}
\end{center}
\end{figure}

The \emph{INTEGRAL}/IBIS imager was useful also in another case, the association of the gamma-ray source 3EG~J$1736-2908$ with the Seyfert 1 galaxy GRS~$1734-292$ ($z=0.0214$, Fig.~\ref{fig:grs1734}, \emph{left panel}), reported by \cite{DICOCCO2004}. The main problem in the association was the radio weakness: although radio observations reported a jet-like structure (Fig.~\ref{fig:grs1734}, \emph{right panel}), the fluxes were of the order of tens mJy \citep{MARTI1998}. A more recent observation at 22~GHz reported an even weaker flux density of $\sim 7$~mJy \citep{MAGNO2025}. It is worth noting that the \emph{Fermi}/LAT catalog confirmed the presence of a gamma-ray source 4FGL~J$1737.1-2901$ consistent with 3EG~J$1736-2908$, but did not confirmed the association with GRS~$1734-292$ \citep{BALLET2023}, which instead was proposed by \cite{MICHIYAMA2024} and \cite{SAKAI2025}. The former found a millimeter excess with Atacama Large Millimeter/submillimeter Array (ALMA) and suggested two possibile explanations: either synchrotron emission from the corona or synchrotron and free-free emission from disk winds \citep{MICHIYAMA2024}. The latter strengthened the disk wind option \citep{SAKAI2025}, a model already used to explain -- at least, partially -- high-energy gamma rays from nearby Seyferts (e.g. \citealt{LENAIN2010,PERETTI2025}). 

Although the disk wind hypothesis sounds well, I would not discard yet the relativistic jet hypothesis, on the basis of the recent discovery of high-frequency radio outburst from radio-quiet/silent NLS1s \citep{LAHTEENMAKI2018}. Follow-up at radio freqiencies showed that these NLS1s have weak or absent radio emission at frequencies below $\lesssim 10$~GHz, likely absorbed by surrounding plasma, but can have strong outbursts at higher frequencies, \citep{BERTON2020,JARVELA2021}. A monitoring campaign with \emph{Swift} on one source of the L\"ahteenm\"aki's sample -- SDSS~J$164100.10+345452.7$ ($z=0.164$) -- found indeed changing absorption at X-rays, with no absorption during the radio outburst and significant absorption during the radio-silent time \citep{ROMANO2023}. In the case of GRS~$1734-292$, almost all the radio observations were done at $\lesssim 10$~GHz, with only two exceptions ($15$ and $22$~GHz, \citealt{MARTI1998,MAGNO2025}) still finding weak emission. However, the discovery of radio outbursts by \cite{LAHTEENMAKI2018} was possible thanks to intensive monitoring, because these bursts are rather erratic and to date we have no idea what possible precursors might be. Therefore, to understand whether GRS~$1734-292$ is a similar case, we need to monitor the source at high radio frequencies to see if it exhibits high-frequency radio bursts. If so, the relativistic jet can easily explain the detection at MeV-GeV energies.

\section{The \emph{Fermi}/LAT gamma-ray sky}
\label{fermisky}
The Large Area Telescope (LAT) onboard the \emph{Fermi Gamma-ray Space Telescope} \citep{ATWOOD2009} has improved the gamma-ray flux limit by about two orders of magnitude ($\sim 10^{-9}$~ph~cm$^{-2}$~s$^{-1}$ vs $\sim 10^{-7}$~ph~cm$^{-2}$~s$^{-1}$) with respect to \emph{CGRO}/EGRET. In addition, LAT has a larger field-of-view ($>2$~sr vs $\sim 0.5$~sr), a better angular resolution and point-source location accuracy ($< 0.5'$ vs $\sim 15'$), a larger energy band ($0.02-300$~GeV vs $0.02-30$~GeV). Already during the first months of operations, it detected high-energy gamma rays from the Seyfert radio galaxy NGC~1275 \citep{ABDO2009A} and the NLS1 PMN~J$0948+0022$ \citep{ABDO2009B}. After the first year of operations, there were four NLS1s \citep{ABDO2009E} and eleven radio galaxies \citep{ABDO2010} plus one (IC~310, \citealt{NERONOV2010}). Among the radio galaxies there are also some sources that were classified as Seyferts, like NGC~1275, NGC~6251, 3C~111, and, particularly, 3C~120 and IC~310, because they are hosted in disk galaxies. There are also the interesting cases of the Seyfert-2 NGC~1068 \citep{LENAIN2010} and the intermediate Seyfert NGC~4151 \citep{PERETTI2025}, whose gamma-ray emission was explained as due partially, or completely, to disk winds. 

The case of NGC~4151 deserves some more words, given the MeV detections in the 70s-80s. According to \cite{PERETTI2025}, there is a nearby contaminating BL Lac Object (1E~$1207.9+3945$, $z=0.617$, which is the nearby source at $\sim 5'$ reported by \citealt{PEROTTI1981A}), but it dominates the gamma-ray emission above $\sim 10$~GeV, while at lower energies, it is the Seyfert to share most of the emission. However, this claim was challenged by \cite{AJELLO2021} and \cite{MURASE2024}. The gamma-ray flux of NGC~4151 estimated by \cite{PERETTI2025} is $\sim 10^{-12}$~erg~cm$^{-2}$~s$^{-1}$ in the $0.1-10$~GeV energy band, at the limit of the LAT performance. The \emph{Fermi}/LAT catalog of point sources detected at $E<100$~MeV does not report any detection starting from 30~MeV after about $8.7$~years of data \citep{PRINCIPE2018}. Although the past detections were at lower energies ($0.2-20$~MeV), it is worth noting that the MeV flux reported by \cite{DICOCCO1977,PEROTTI1983,BAITY1984} was quite high, of the order of $\sim 10^{-8}$~erg~cm$^{-2}$~s$^{-1}$ in the $1-20$~MeV energy range, of the same order of magnitude of the flux above 100~MeV reported by \cite{GALPER1979}. The strong variability might still be an option as suggested by \cite{PINKAU1980}, \cite{JOURDAIN1992} and \cite{MAISACK1993}. Once in a lifetime fluxes have been recorded from other sources with more robust multiwavelength observations: for example, the 2009 big outburst from 3C~454.3 with a $0.1-100$~GeV flux of $\sim 6\times 10^{-9}$~erg~cm$^{-2}$~s$^{-1}$ \citep{BONNOLI2011}. Therefore, the gamma-ray detection of NGC~4151 in the 1970s and 1980s could also fall into these rare cases.

\cite{FOSCHINI2021,FOSCHINI2022} revised the classification and redshift of 2980 extragalactic gamma-ray sources of the 4FGL data release 2, by searching in all the available literature. Table~\ref{tab:4fglrev} displays the new classification updated with the recent optical spectra from \cite{RAJAGOPAL2023}, \cite{GARCIAPEREZ2023}, and \cite{DALLABARBA2026}.

\begin{table}[t]
\caption{Revised classification of 2980 extragalactic gamma-ray point sources from the 4FGL-DR2 according to \cite{FOSCHINI2021,FOSCHINI2022}. Columns: (1) revised class (BLLAC: BL Lac Object; FSRQ: flat-spectrum radio quasar; NLS1: narrow-line Seyfert 1 galaxy/quasar; SEY: Seyfert galaxy; MIS: misaligned AGN; UNCL: unclassified; AMB: ambiguous; CLAGN: changing-look AGN); (2) number of objects; (3) 4FGL-DR2 class (BLL: BL Lac Object; FSRQ: flat-spectrum radio quasar; NLSy1: narrow-line Seyfert 1 galaxy/quasar; SEY: Seyfert galaxy; RDG: radio galaxy; SSRQ: steep-spectrum radio quasar; CSS: compact steep spectrum quasar; BCU: active galaxy of uncertain type; UNK: unknown; AGN: other non-blazar active galaxy; the difference between uppercase and lowercase refers to the significance of the identification); (4) number of objects; (5) difference of objects between the revised classification and the 4FGL-DR2; (6) percent of objects with spectroscopic redshift. Note that the 10 objects generically classified as AGN/agn in the 4FGL-DR2 have been reclassified as BLLAC, MIS, CLAGN, and UNCL and moved to their respective classes. }
\begin{center}
\begin{tabular}{lclccc}
\hline
\multicolumn{2}{c}{\centering {\bf rev4FGL}} & \multicolumn{2}{c}{{\bf 4FGL-DR2}} & {\bf Difference} & {\bf Spectroscopic $z$}\\
(1) & (2) & (3) & (4) & (5) & (6)\\
\hline
BLLAC & 1228 & BLL/bll & 1204 & $+24$ & 47.2\%\\
FSRQ  & 692 & FSRQ/fsrq & 703 & $-11$ & 99.7\%\\
NLS1 & 25 & NLSy1/nlsy1 & 9 & $+16$ & 100\%\\
SEY & 35 & SEY/sey & 0 & $+35$ & 100\%\\
MIS & 85 & RDG/rdg & 45 & $+40$ & 96.5\%\\
{} & {} & SSRQ/ssrq & {} & {} & {}\\
{} & {} & CSS/css & {} & {} & {}\\
UNCL & 839 & BCU/bcu & 1009 & $-170$ & 0.1\%\\
{} & {} & UNK/unk & {} & {} & {}\\
AMB & 41 & {} & {} & {} & 70.7\%\\
CLAGN & 35 & {} & {} & {} & 100\%\\
\multicolumn{2}{l}{1 BLLAC} & AGN/agn & 10 & {} & {}\\
\multicolumn{2}{l}{3 MIS} & {} & {} & {} & {}\\
\multicolumn{2}{l}{3 CLAGN} & {} & {} & {} & {}\\
\multicolumn{2}{l}{3 UNCL} & {} & {} & {} & {}\\
\hline
\end{tabular}
\end{center}
\label{tab:4fglrev}
\end{table}

Two updates are worth mentioning. \cite{ZENG2025} noted the unusual radio morphology of the FSRQ $0954+556$ a.k.a. 4FGL~J$0957.6+5523$, with significantly different jet directions at kpc and pc scales, and hypothesized that there may have been a reorientation of the jet axis. This might be the second case of CLAGN with jet reorientation in the present sample of gamma-ray emitting jetted AGN, the other being 4FGL~J$2334.9-2346$ (PKS~$2331-240$, \citealt{HERNANDEZ2017}).

Another case worth noting is 4FGL~J$1154.0+4037 = $B$3$~$1151+408$ ($z=0.9251$). The first redshift was measured by \cite{HOOK1996} on the basis of one emission line identified as Mg~II. \cite{HENSTOCK1997} confirmed the redshift and, with a better coverage of red wavelengths ($\lambda_{\rm obs}\sim 9358$\AA\, for H$\beta$), detected also the H$\beta$-[OIII] complex and measured the full-width half-maximum (FWHM) of the lines. They found the surprising result of FWHM(Mg~II)$\sim 8900$~km~s$^{-1}$ and  FWHM(H$\beta$)$\sim 1900$~km~s$^{-1}$. The \href{https://skyserver.sdss.org/DR19/VisualTools/explore/summary?ra=178.477745&dec=40.614616}{Sloan Digital Sky Survey (SDSS) DR19} display a spectrum with a prominent Mg~II line, but the H$\beta$-[OIII] is again buried into the noise. Even the new optical spectrum taken by \cite{DALLABARBA2026} does not solve the issue, as it has the same noise problem at red wavelengths, but it confirms the huge width of the Mg~II line, with FWHM(Mg~II)$\sim 10100$~km~s$^{-1}$. A new optical spectrum with high-quality coverage of red-infrared wavelengths is necessary to understand the unusual nature of this object.

\begin{figure}[!t]
\begin{center}
\includegraphics[scale=0.15]{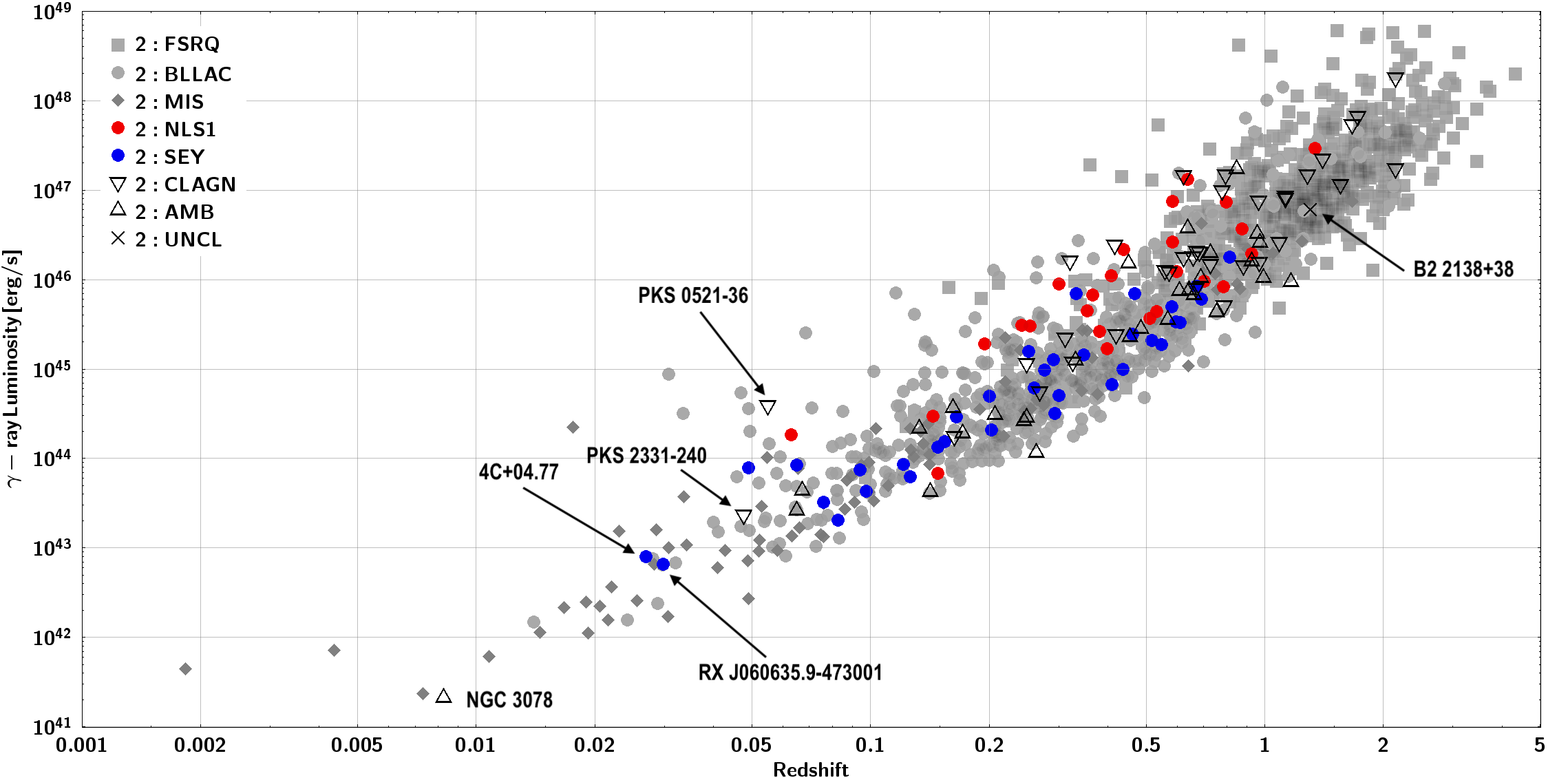}
\caption{Gamma-ray luminosity vs redshift of the 1477 objects with spectroscopic redshifts.}
\label{fig:gammaz}
\end{center}
\end{figure}

\begin{figure}[!t]
\begin{center}
\includegraphics[scale=0.15]{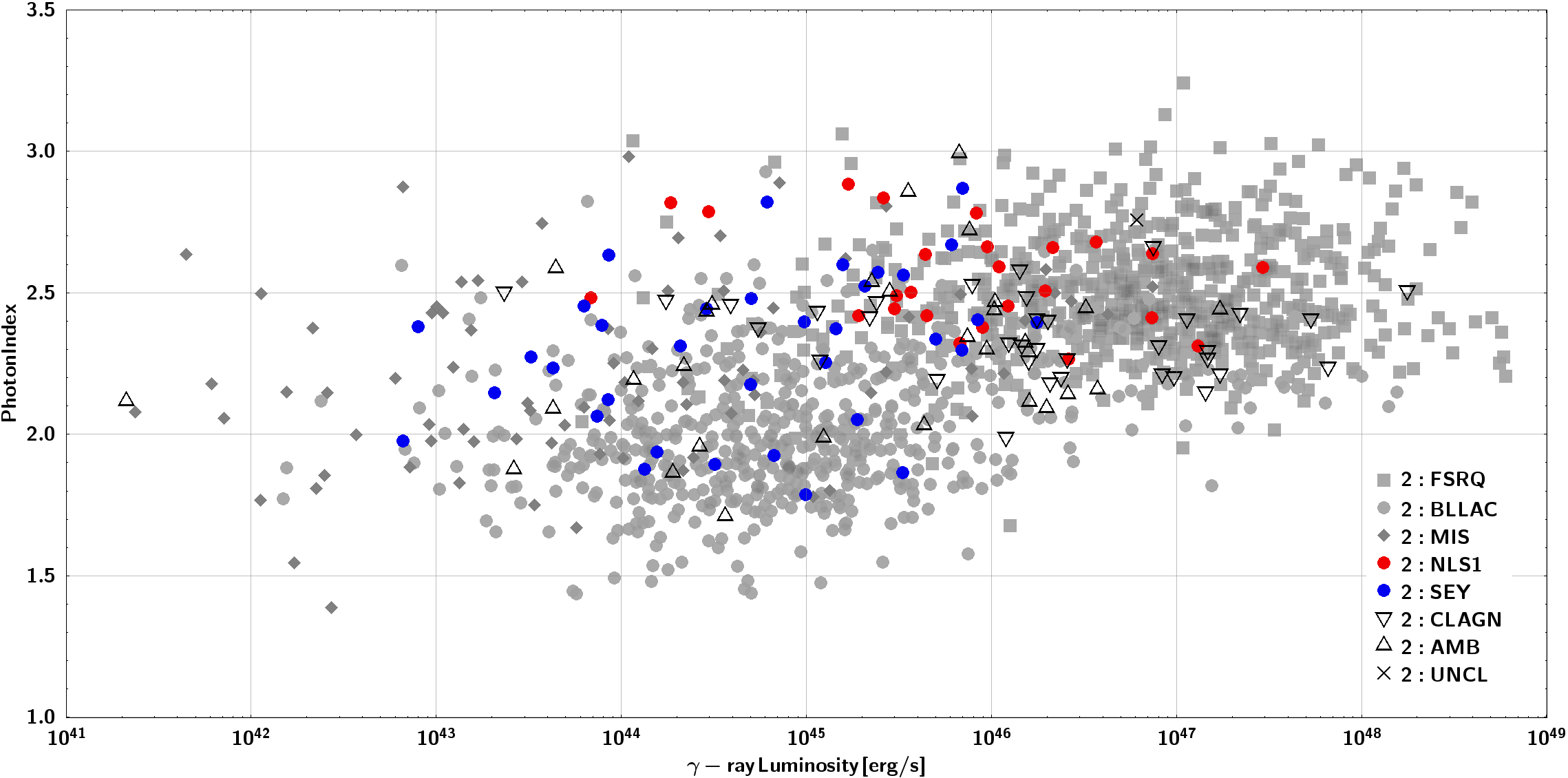}
\caption{Gamma-ray luminosity vs photon index of the 1477 objects with spectroscopic redshifts.}
\label{fig:gammagamma}
\end{center}
\end{figure}

\section{Gamma-ray properties}
The rev4FGL catalog has 1477/2980 spectroscopic redshifts. These 1477 objects are divided into: 690 FSRQ, 580 BLLAC, 82 MAGN, 25 NLS1, 35 SEY, 35 CLAGN, 29 AMB, and 1 UNCL. Looking at the Fig.~\ref{fig:gammaz}, given the position of the only UNCL (B2~$2138+38$), it is likely to be classified as FSRQ. The CLAGN and AMB are randomly distributed along the stream of blazars, as somehow expected. Particularly, CLAGN in this case often means a change from line-dominated to featureless optical spectrum and vice versa, i.e. a change from FSRQ to BLLAC and the opposite, likely depending on the jet activity rather than a change in the accretion rate. MIS are distributed at low redshift and luminosity. One AMB (NGC~3078) and a few SEY are in that zone, and therefore might be an indication of a MIS nature. The two SEY are: RX~J$060635.9-473001$, classified as LINER/Seyfert 2 by \cite{PIETSCH1998}, and 4C~$+04.77$, originally classified as BLLAC, but \cite{VERON1993} showed that the active nucleus is a type-1 Seyfert, whose emission lines are overwhelmed by the light of the giant elliptical host galaxy. 

Some CLAGN are also MIS (for example, there is the case of PKS~$0521-36$, already noted by Woltjer in \citealt{BLANDFORD1990}), but this time the change of the optical spectrum might be due to changes in the accretion, as the jet viewed at large angles is unlikely to overwhelm the emission lines during outbursts. It is also worth reminding the already cited case of PKS~$2331-240$: \cite{HERNANDEZ2017} suggested that it was a radio galaxy, whose jet had changed direction in recent times and is now pointing towards Earth. The optical spectrum displays the emission lines of the complexes H$\alpha$-[NII] and H$\beta$-[OIII], with weak H$\beta$, and \cite{HERNANDEZ2017} proposed a type-1.9 Seyfert classification. This might be consistent with a large viewing angle and being placed in the low-power region among BLLAC and MIS, while the inverted radio spectrum of the core suggests that the jet is pointing toward the Earth. Interestingly, the appearance of a bump in the $\sim 6500-6900$\AA\, region of the 2019 spectrum, might be due to a reprocessing of the broad-line region and the presence of dust-driven winds (compare with \citealt{JAISVAL2025}), which favor the proposed classification as Seyfert 1.9. 

\begin{figure}[!t]
\begin{center}
\includegraphics[scale=0.15]{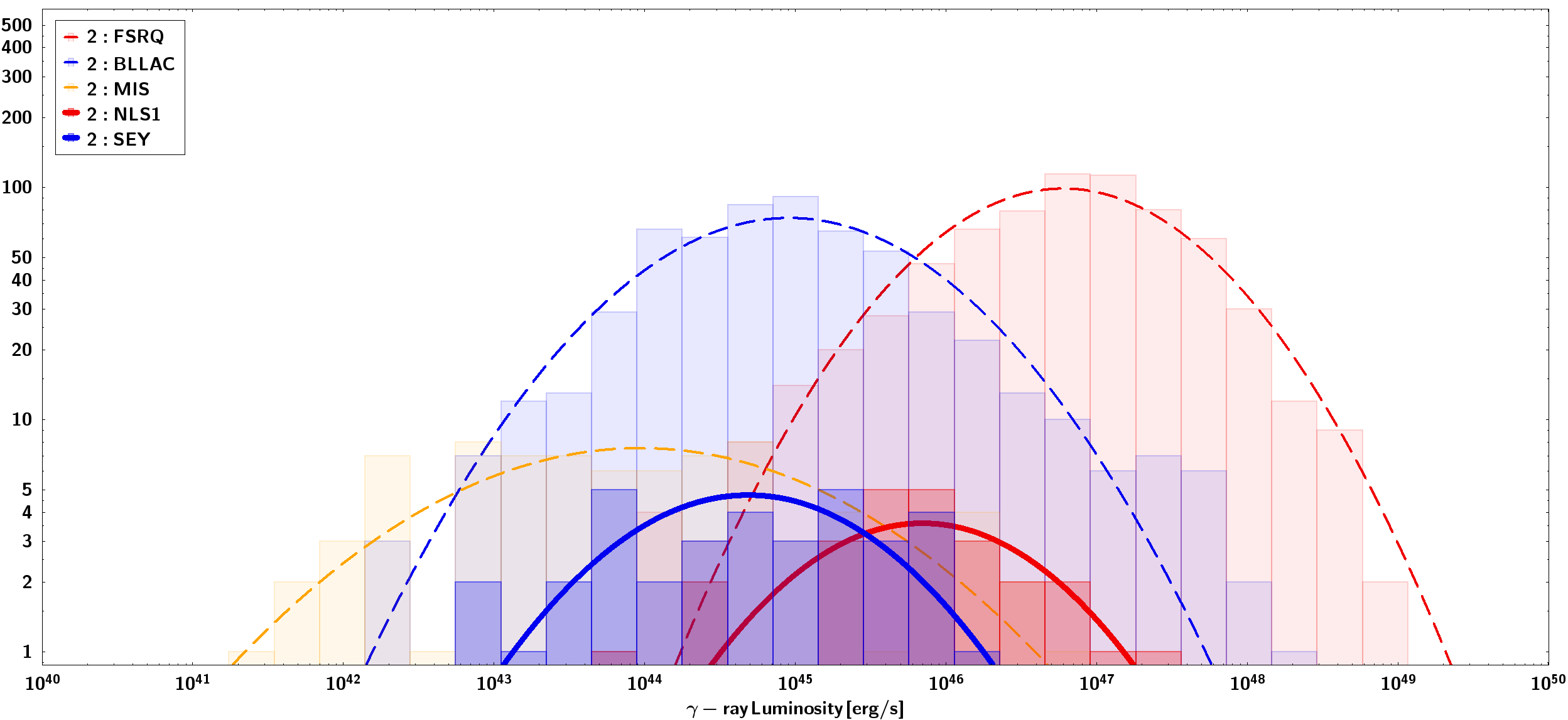}
\caption{Distribution of the gamma-ray luminosity of the jetted AGN with spectroscopic redshift, excluding CLAGN, AMB, and UNCL objects.}
\label{fig:gammapow}
\end{center}
\end{figure}

Fig.~\ref{fig:gammagamma} displays the gamma-ray luminosity vs the photon index. Again, the UNCL B2~$2138+38$ is in the region populated by FSRQs. Interestingly, NLS1-SEY populations appear to follow the same trend as FSRQ-BLLAC populations. 

Fig.~\ref{fig:gammapow} shows the distribution of the gamma-ray luminosity of the objects with certain classification, i.e. after having removed the CLAGN, AMB, and UNCL sources. As expected, the NLS1s are the low-luminosity tail of the FSRQs distribution \citep{BERTON2016}, while SEY seems to be something between the BLLAC distribution and the high-luminosity tail of MIS. This might be due to the fact that the SEY class is a mixture of type-1, intermediate, and type-2 Seyferts, thus jets with a variety of viewing angles. 

\section{Cross-match with the MOJAVE radio catalog}
Monitoring Of Jets in Active galactic nuclei with VLBA Experiments (MOJAVE\footnote{\url{https://www.cv.nrao.edu/MOJAVE/}}) is a long-term program of radio observations at 15~GHz of jetted AGN \citep{LISTER2018}. It offers a wealth of astronomical data. Particularly, \cite{HOMAN2021} measured and calculated the brightness temperature, synchrotron peak frequency, apparent jet speed $\beta$ [in units of $c$, speed of light in vacuum], bulk Lorentz factor $\Gamma$, Doppler factor $\delta$, and viewing angle $\theta$ [deg or rad] for 447 jetted AGN, based on radio observations from 1994 to 2019. A list of 282 objects is obtained by cross-matching the MOJAVE data with the current rev4FGL sample of 1477 gamma-ray AGN with spectroscopic redshifts. The list is divided as follows: 173 FSRQs, 66 BLLAC, 14 MIS, 7 NLS1s, 4 SEY, 15 CLAGN, 3 AMB. From these data, it is possible to derive further quantities, such as the jet opening angle $\phi$ [deg or rad] and the jet power according to \cite{BLANDFORD1979}. Since there are some slight differences in the redshift values used by \cite{HOMAN2021} and the values we reported in the rev4FGL \citep{FOSCHINI2021,FOSCHINI2022}, I have corrected the affected quantities by following the procedures described in \cite{FOSCHINI2024}.

\begin{figure}[!t]
\begin{center}
\includegraphics[scale=0.13]{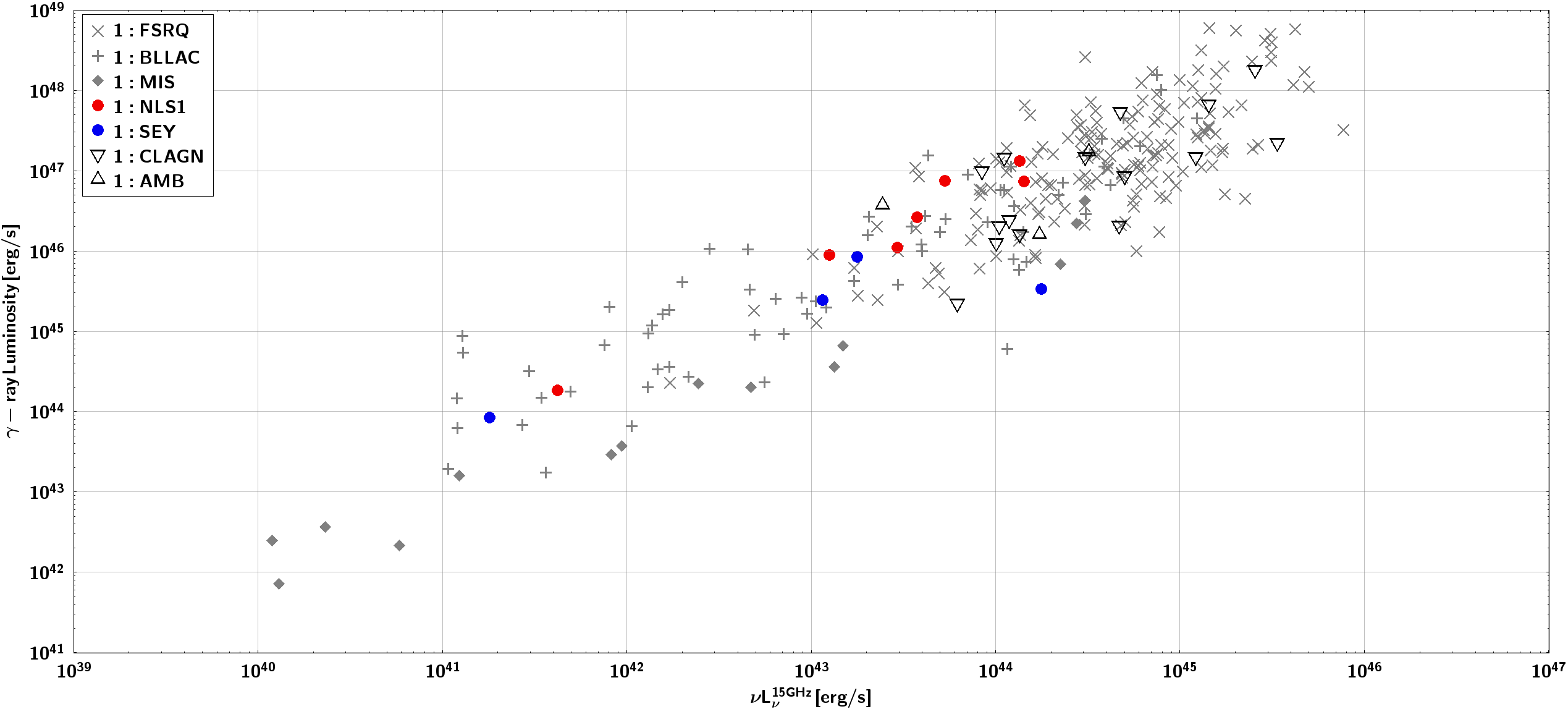}\\
\includegraphics[scale=0.13]{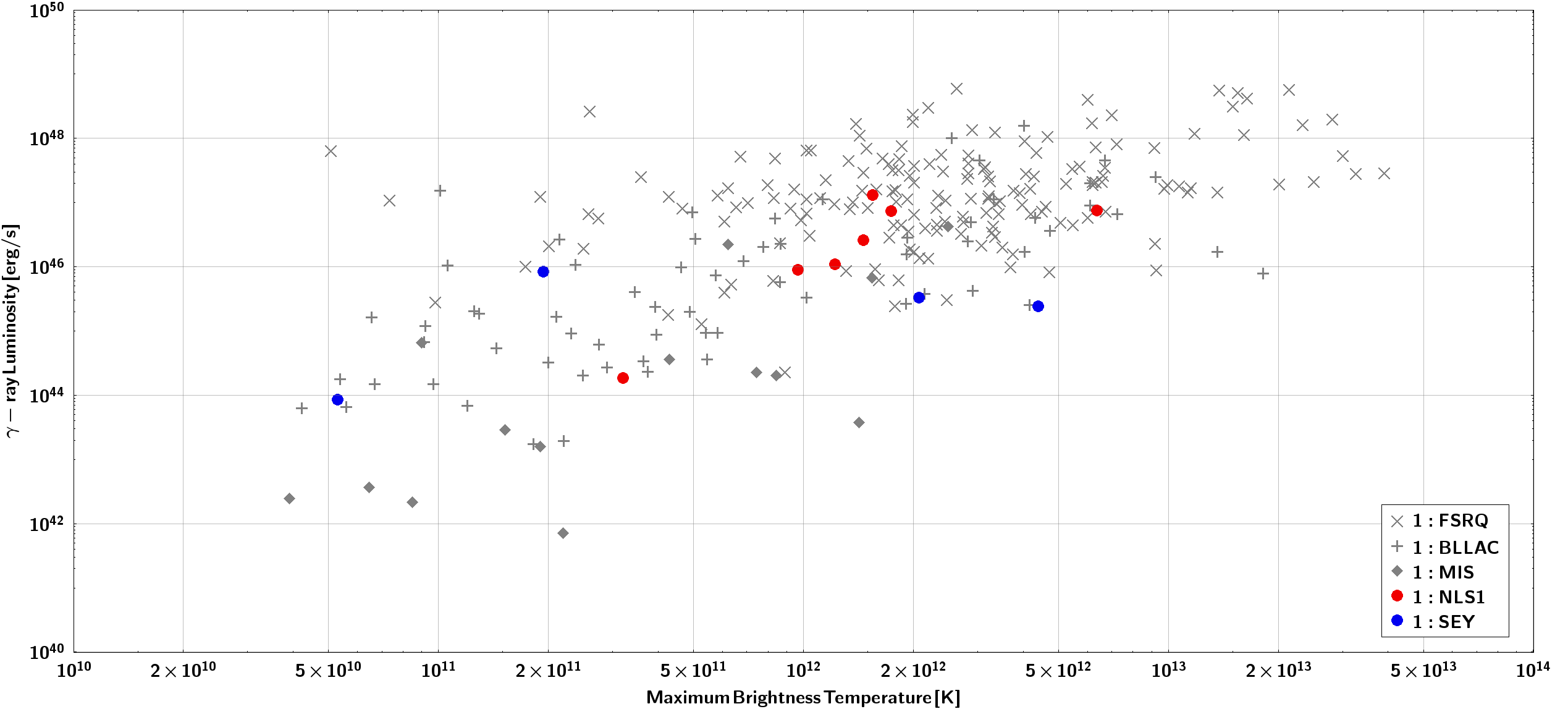}
\caption{Gamma-ray luminosity versus: (\emph{top panel}) radio luminosity at 15~GHz, and (\emph{bottom panel}) maximum brightness temperature.}
\label{fig:gammaradio}
\end{center}
\end{figure}

Fig.~\ref{fig:gammaradio}, \emph{top panel}, displays the gamma-ray versus the radio luminosity ($L_{\gamma}$ vs $L_{\rm 15\,GHz}$), confirming the well-known relationship (e.g. \citealt{LISTER2011,GHIRLANDA2011}). The other panel show the gamma-ray luminosity versus the maximum brightness temperature $T_{\rm b,max}$. There might be a trend, from low-luminosity and low $T_{\rm b,max}$ of radio galaxies to the high-luminosity and high $T_{\rm b,max}$ of FSRQs, but the latter class is distributed as well also at high-luminosity and low $T_{\rm b,max}$. This is likely due to the angular resolution problems in the VLBA radio observations, and affects more high-$z$ objects as FSRQs (\citealt{HOMAN2021}; cf also \citealt{FOSCHINI2024}).

\begin{figure}[!t]
\begin{center}
\includegraphics[scale=0.09]{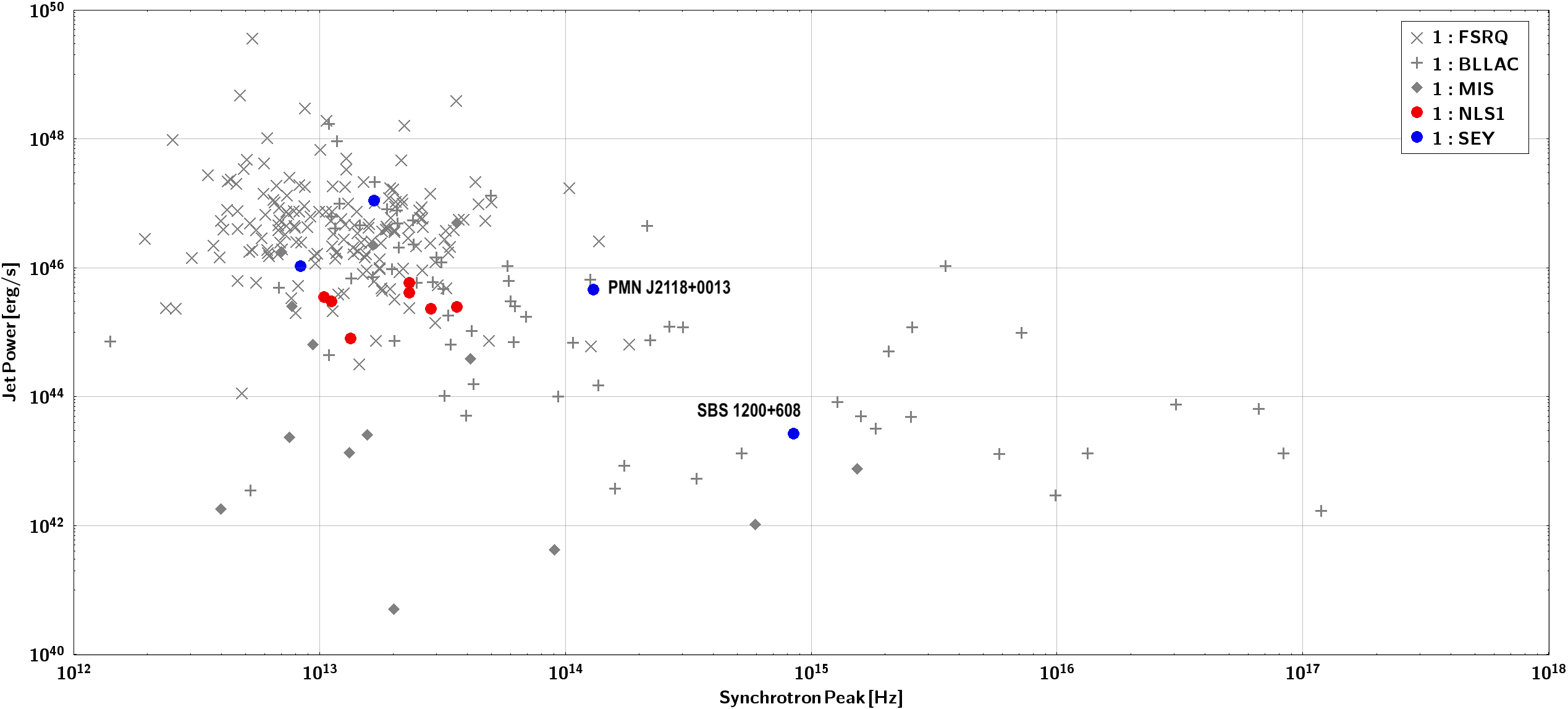}
\includegraphics[scale=0.09]{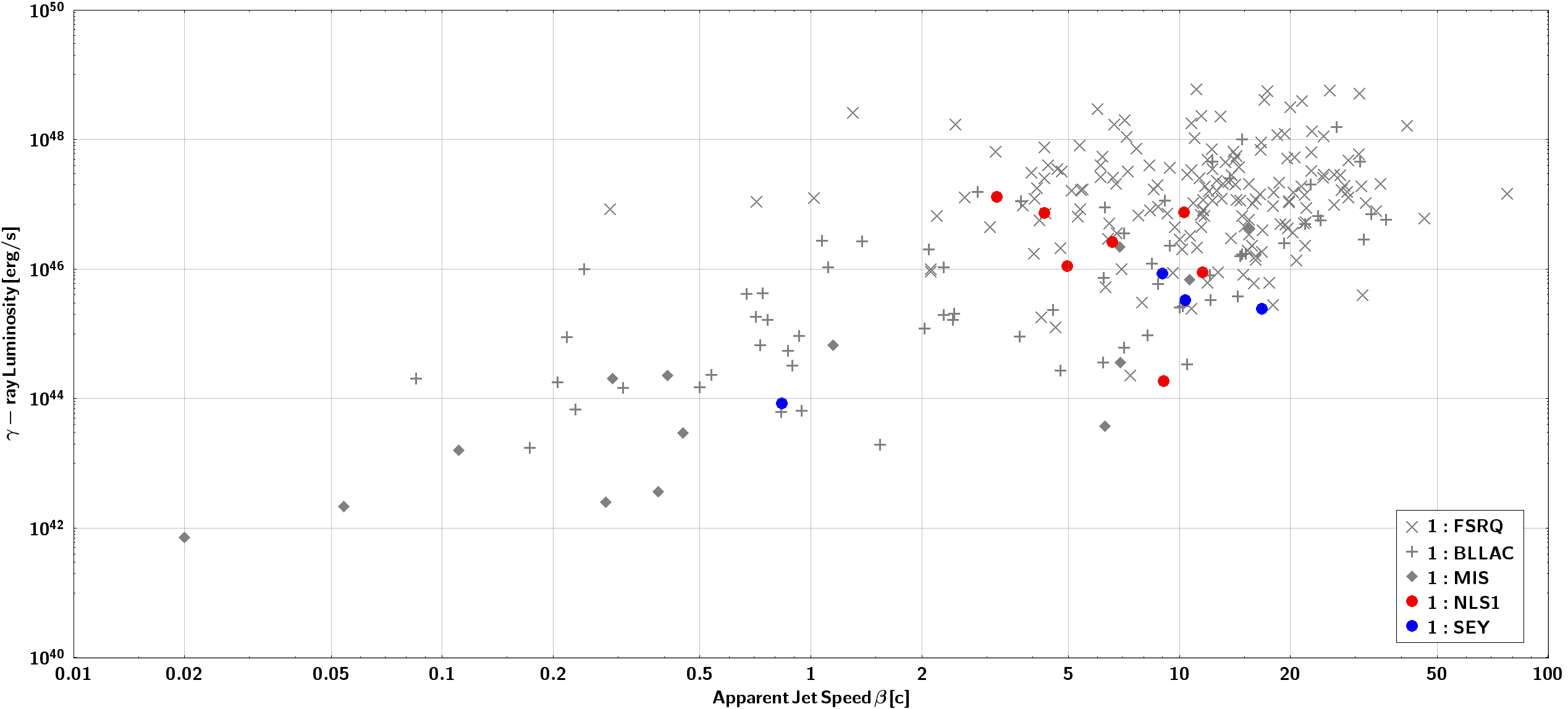}\\

\includegraphics[scale=0.09]{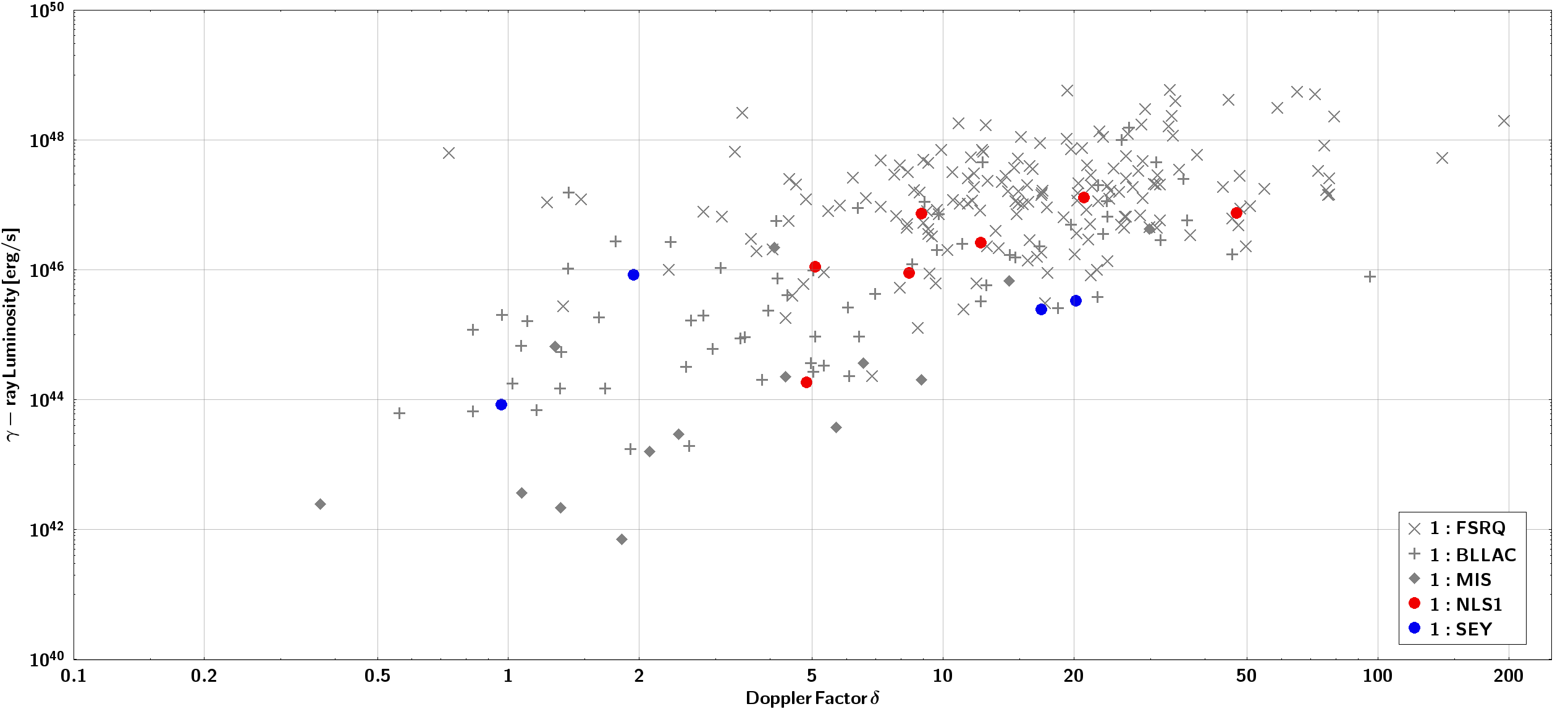}
\includegraphics[scale=0.09]{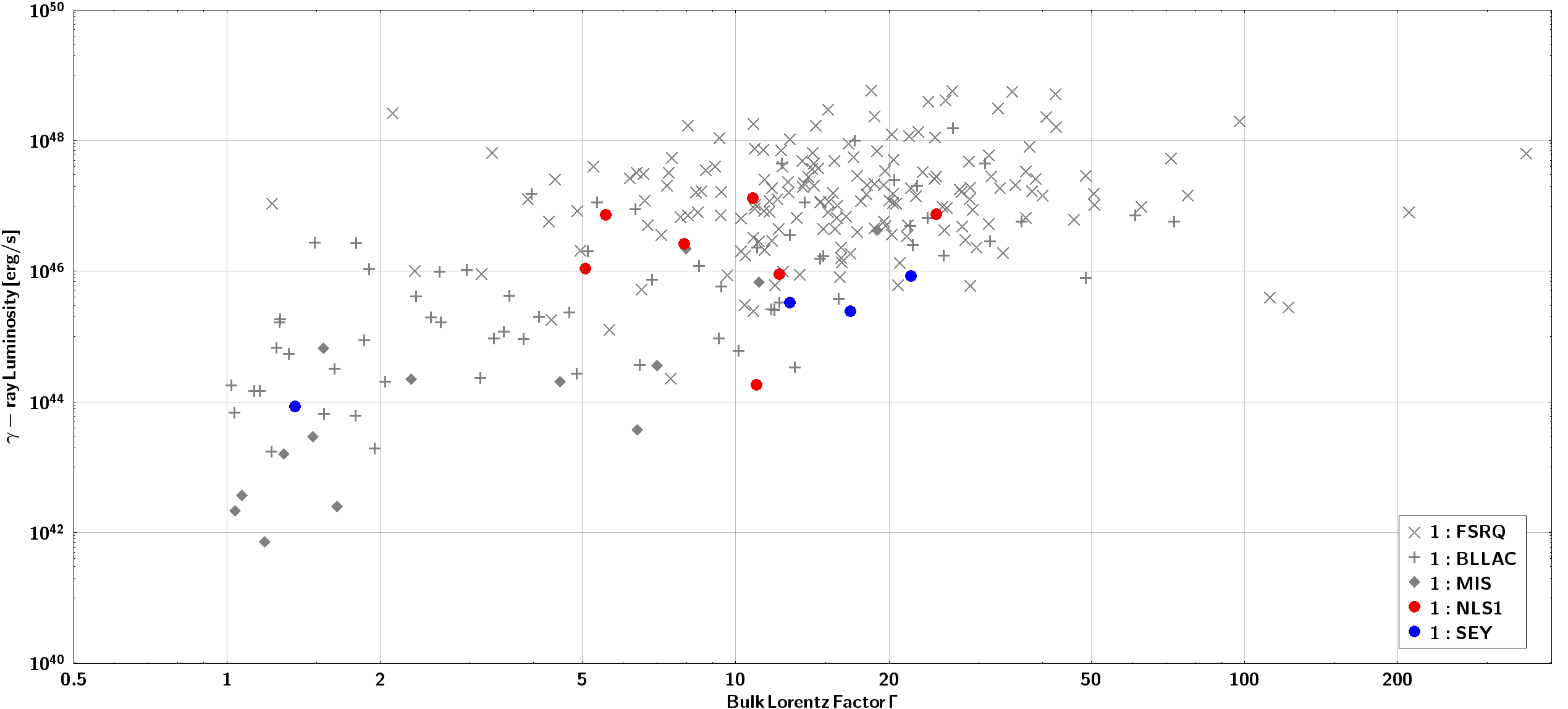}\\

\includegraphics[scale=0.09]{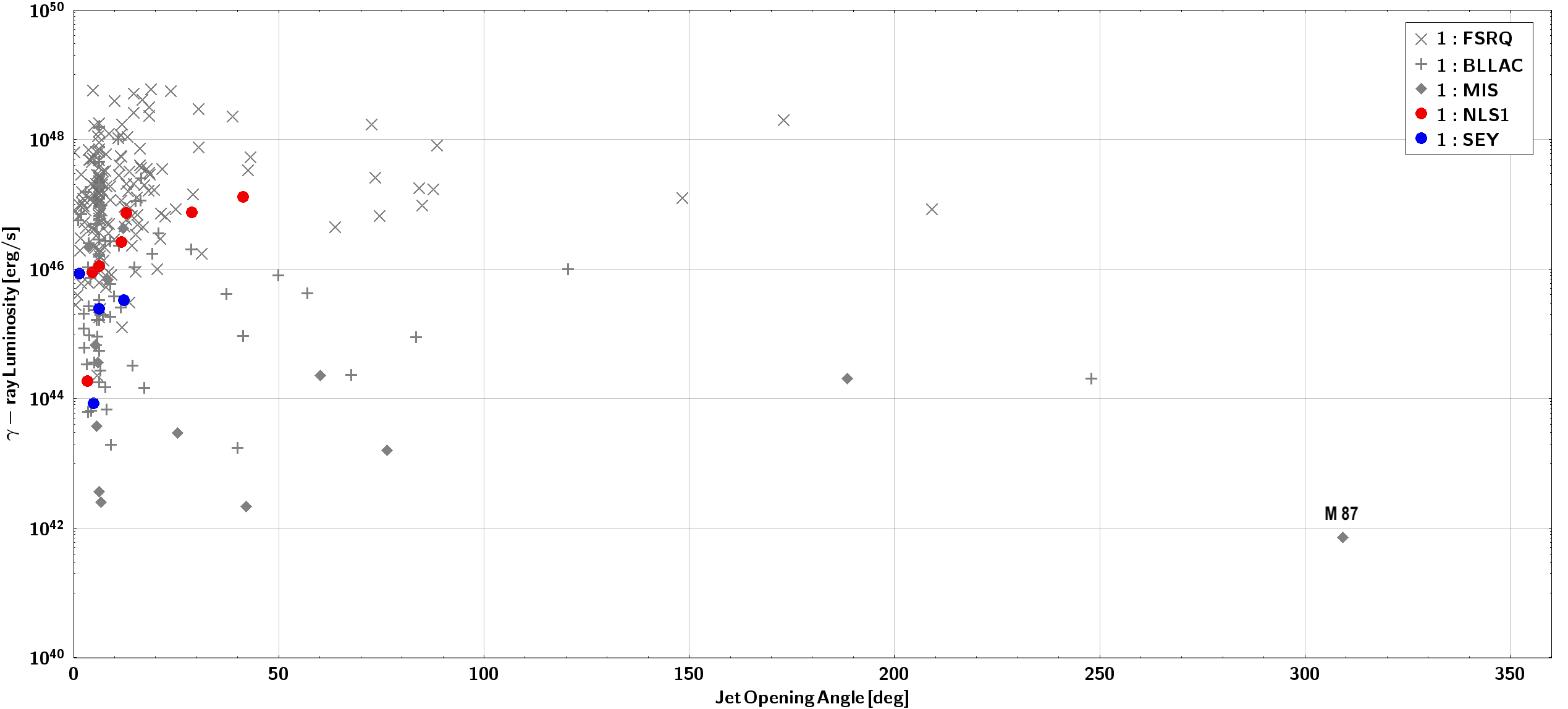}
\includegraphics[scale=0.09]{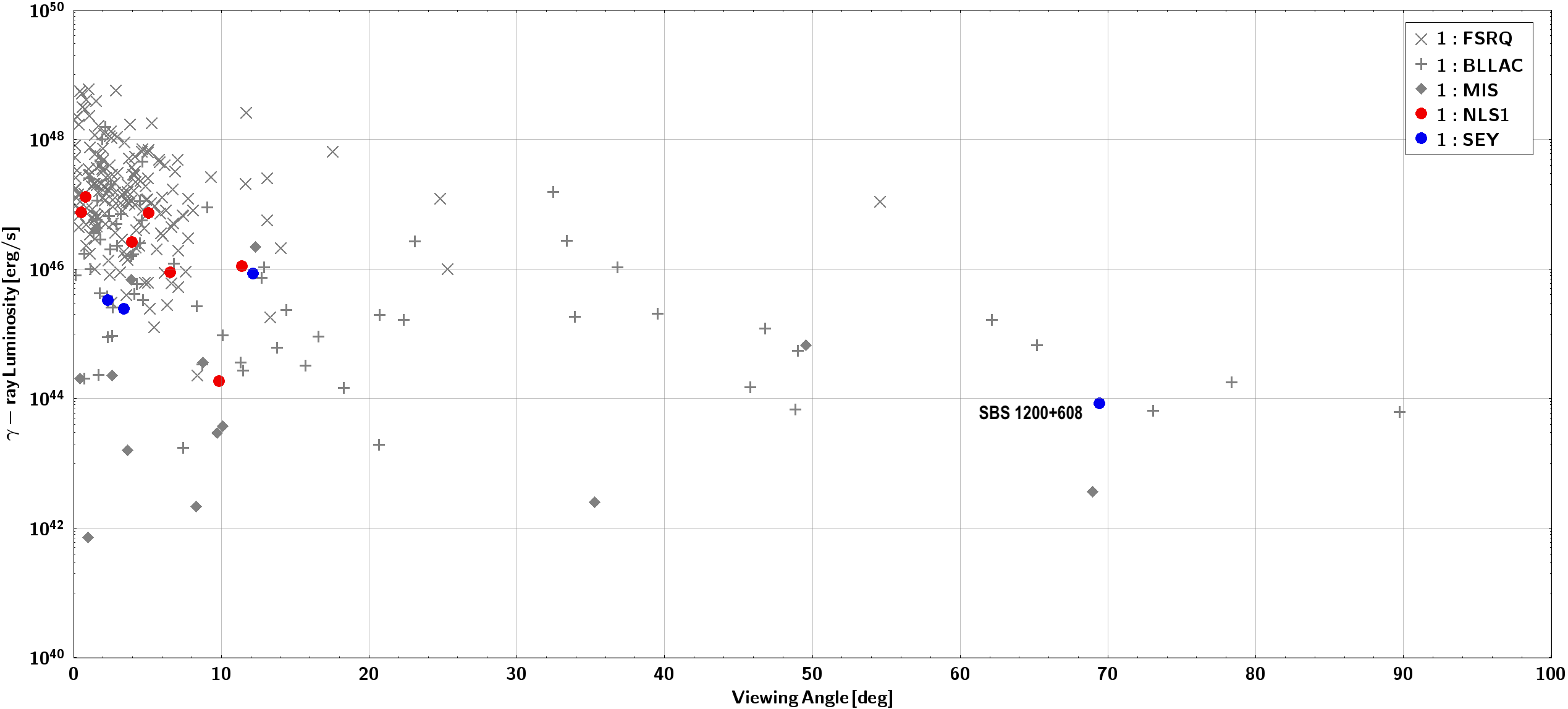}\\

\caption{(\emph{top left panel}) Total jet power vs synchrotron peak frequency. All the other panels display the gamma-ray luminosity versus: (\emph{top right panel}) apparent jet speed $\beta$, (\emph{middle left panel}) Doppler factor, (\emph{middle right panel}) bulk Lorentz factor, (\emph{bottom left panel}) jet opening angle, (\emph{bottom right panel}) viewing angle.}
\label{fig:mojave}
\end{center}
\end{figure}

More comparisons between different quantities measured or calculated from the MOJAVE radio observations are shown in Fig.~\ref{fig:mojave}. Particularly, the \emph{top left panel} shows $P_{\rm jet}$ versus the synchrotron peak frequency, which is one way to display the so-called blazar sequence \citep{FOSSATI1998}. Although we have a small number of NLS1-SEY, it is possible to see that NLS1s are localized on the border between FSRQ and BLLAC, while the four SEY are divided: two in the FSRQ region and two in the BLLAC zone. These two latter are SBS~$1200+608$ ($z=0.065$) and PMN~J$2118+0013$ ($z=0.463$). The former was classified as Seyfert in the Second Byurakan Spectral Sky Survey (SBS), but \cite{MARTEL1994} reclassified it as LINER hosted in a S0 galaxy. The \href{https://skyserver.sdss.org/DR19/VisualTools/explore/summary?ra=180.7646&dec=60.522}{SDSS spectrum} displays an almost featureless noisy spectrum with a prominent H$\alpha$-[NII] complex. The 4FGL classified it as BLLAC, but the equivalent width of the H$\alpha$ is surely greater than the standard threshold of 5\,\AA. It is interesting to note that this object also shows a large viewing angle ($\sim 70^{\circ}$, see Fig.~\ref{fig:mojave}, \emph{bottom right panel}), thus suggesting a type-1.9 Seyfert optical classification and MIS radio classification (indeed, MIS are distributed in the low-power zone of the plot). 

The other source, PMN~J$2118+0013$, is a Seyfert intermediate according to \cite{MASSARO2014}, although other analyses suggest a NLS1 classification (e.g. \citealt{RAKSHIT2020}). The \href{https://skyserver.sdss.org/DR19/VisualTools/explore/summary?ra=319.5725&dec=0.2213}{SDSS spectrum} seems to favor the Seyfert type-1.5 classification. 

The other panels of Fig.~\ref{fig:mojave} confirm the analogous plots published by \cite{HOMAN2021}, with the tendency to saturate at high luminosities for $\delta$ and $\Gamma$ (\emph{center panels}), likely because of the already cited dependence on the brightness temperature. 

\section{Do all relativistic jets emit gamma rays?}
Although \emph{Fermi}/LAT detected most of the jetted AGN already detected by \emph{CGRO}/EGRET, not all the 3EG gamma-ray sources were confirmed. Already in the list of 132 high-significance gamma-ray jetted AGN detected after the first three months of \emph{Fermi}/LAT operations, there were only 35 of them also with an EGRET detection \citep{ABDO2009C}. \cite{FOSCHINI2022} made a cross-match of the rev4FGL with the 3EG and found only 100 matches over 183 EGRET sources\footnote{The rev4FGL excluded the gamma-ray sources with Galactic latitude $|b|<10^{\circ}$ to avoid the problems of the diffuse gamma-ray emission on the Galactic plane, and contained 2980 extragalactic gamma-ray sources with a counterpart. The third EGRET catalog (3EG) contains 271 gamma-ray sources, but only 183 are outside the Galactic plane. I excluded the pulsar CTA~1 (3EG~J$0010+7309$), the Large Magellanic Cloud (3EG~J$J0533-6916$), and $\rho$ Oph (3EG~J$1627-2419$), which have $|b|>10^{\circ}$.}. The obvious question is whether the 83 gamma-ray sources in the EGRET catalog are missing in the rev4FGL because of source variability or there are other reasons. Moreover, since we were interested in NLS1s, we checked the three cases involving this class of AGN: in two of these cases, we found that, although the \emph{Fermi}/LAT detection was consistent with the \emph{CGRO}/EGRET one, the counterpart was different \citep{FOSCHINI2022}. Again, one immediately wonders whether this change is due only to an improvement in point-source location accuracy or whether the jetted AGN, which was emitting gamma rays at the time of EGRET, has now turned off, while another jetted AGN, which was off at the time of EGRET, has now turned on. 

Comparison of the radio properties between gamma-ray detected and non-detected jetted AGN were done after \emph{CGRO}/EGRET results. It was found that gamma-ray jetted AGN are more compact on parsec scale, more luminous, more polarized, have faster apparent jet speed, higher core dominance, and higher brightness temperature, than non-gamma jetted AGN (\citealt{KELLERMANN2004,KOVALEV2005,LISTER2005,TAYLOR2007}). However, \emph{CGRO}/EGRET was limited both in the field-of-view and flux, making it necessary to point the source in outburst, and therefore it was not possible to make a systematic comparison between radio and gamma-ray properties of jetted AGN. 

These limitations were overcome with \emph{Fermi}/LAT, which improved sensitivity by two orders of magnitude and the large field of view allowed the sky to be scanned every three hours until a fairly homogeneous coverage was obtained. New studies confirmed the above scenario: gamma-ray jetted AGN have ``more'' of some quantity than non-gamma jetted AGN (\citealt{KOVALEV2009,LISTER2009,PUSHKAREV2009,SAVOLAINEN2010,LINFORD2011,LISTER2011,WU2014,XIAO2019}). On the basis of these radio properties, \cite{LISTER2015} proposed three non-gamma jetted AGN to be potential candidates for gamma-ray detection: they are III~Zw~2 (a.k.a. Mrk~$1501$, an intermediate Seyfert), PKS~$0119+11$, and 4C~$+69.21$. The two latter were indeed inserted in the 4FGL-DR4 \citep{BALLET2023}, while Mrk~$1501$ is not present (yet?), although there was a claim of gamma-ray detection (\citealt{LIAO2016}, see also \citealt{CHEN2010}). \cite{LIAO2016} analyzed the first seven years of \emph{Fermi}/LAT data, and found a significant detection ($\sim 6.2\sigma$) only in the second year, plus some other short flares on hourly time scales, and weak detections. This would explain why the source is not in the 4FGL: one or a few detections on short timescales are lost when integrated over a long timescale (14 years for the 4FGL-DR4). Indeed, Liao's findings were confirmed by the \emph{Fermi}/LAT catalog of transient sources (1FTL, \citealt{BALDINI2021}), where the search for detections was done on monthly bins. 

It is evident a link between radio properties and gamma-ray emission in jetted AGN: there is a well-grounded physical theory, based on synchrotron and inverse-Compton processes, plus a significant correlation between observed data. However, something is still missing. As already noted by \cite{JORSTAD2001}, although most gamma-ray flares are correlated with the ejection of superluminal radio components, there is not a one-to-one correlation (see also the review \citealt{JORSTAD2016}). There are also the so-called ``orphan flares'', gamma-ray flares without counterpart at lower frequencies \citep{KRAWCZYNSKI2004}, and ``sterile flares'', optical flares without gamma-ray detection \citep{RAITERI2025}, and also more recent multiwavelength campaigns are showing intriguing deviations from the standard scenario (e.g. \citealt{SHIAO2025}; see also \citealt{RAITERI2025} for an extended review on variability). 

\begin{figure}[!t]
\begin{center}
\includegraphics[scale=0.3]{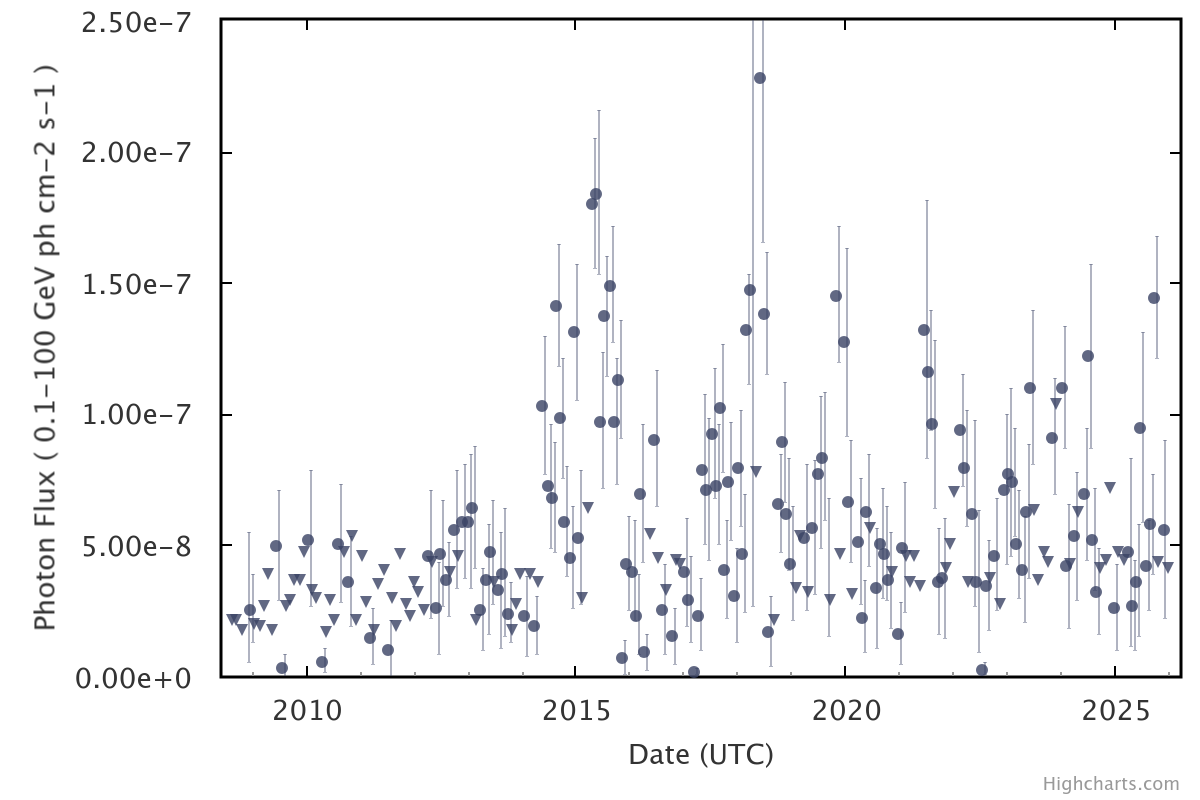}
\caption{\emph{Fermi}/LAT lightcurve of the Seyfert galaxy 3C~120 (monthly bins, $0.1-100$~GeV, from the \emph{Fermi}/LAT Lightcurve Repository,  \citealt{ABDOLLAHI2023}).}
\label{fig:3c120}
\end{center}
\end{figure}

Changes in the jet structure and morphology can explain some variability in the gamma-ray emission. As discussed in Sect.~\ref{fermisky}, the rev4FGL contains two candidates cases of significant change in the jet morphology, moving from large to small viewing angles \citep{HERNANDEZ2017,ZENG2025}, but there are also cases of less dramatic and periodic changes, such as helical jets (e.g. \citealt{VILLATA1999,CAPRONI2004,RAITERI2025}). One work is worth noting: \cite{CASADIO2015} explained the gamma-ray emission from the Seyfert galaxy 3C~120 as the change of viewing angle -- from $9^{\circ}.2$ to $3^{\circ}.6$ -- induced by the helicoidal motion of the plasma, according to the model by \cite{CAPRONI2004}, although the phases do not match. Indeed, the gamma-ray lightcurve (Fig.~\ref{fig:3c120}) does not show any evidence of periodicity. As noted by \cite{CASADIO2015}, 3C~120 started its gamma-ray activity in 2012, four years after the launch of \emph{Fermi}, and had other outbursts unevenly separated after the Casadio's work, while the helical model by \cite{CAPRONI2004} required a $12.4$ years periodicity. Anyway, the observation of a change of the inner jet viewing angle is inspiring to study the physical mechanism at work, and can give an interesting and simple explanation of the gamma-ray emission of misaligned AGN.

Structural differences of the jet were already proposed by \cite{GRANDI2012} and \cite{GRANDITOR2012} to explain the small number of Fanaroff-Riley II (FRII) radio galaxies detected by \emph{Fermi}/LAT with respect to the FRI. As known, the seminal work by \cite{FANAROFF1974} divided the jet morphology of radio galaxies into two classes (FRI, FRII), depending on the jet power. One would expect that FRII, the high-power class, should be easier to be detected at gamma rays than FRI, the low-power class. However, \cite{GRANDI2012} found the opposite, and this is confirmed also in the rev4FGL. Indeed, the 85 gamma-ray sources classified as MIS can be divided into 8 FR0, 35 FRI, 18 FRII, 7 compact steep-spectrum source (CSS), 1 steep-spectrum radio quasar (SSRQ), and 16 with no specified class \citep{FOSCHINI2021,FOSCHINI2022}. \cite{GRANDI2012} proved that the small number of gamma-ray detected FRII is not due to the redshift of the sources of this class, which is greater than that of FRI. The reason could be different jet structure: for example, FRII jets may be less structured, with a fast spine and a few shocked region, while FRI could be characterized by a more pronounced structures, affected by a more efficient deceleration \citep{GRANDI2012,GRANDITOR2012} . \cite{ANGIONI2020} studied a sample of misaligned AGN and found that the gamma-ray emission is driven by the inner parsec-scale jet rather than the core brightness and dominance, as happened in the case of 3C~120 \citep{CASADIO2015}.

Another hypothesis to explain detection or not at gamma rays of FSRQs was put forward by \cite{DENG2021}: they cross-matched the Candidate Gamma-Ray Blazar Survey (CGRaBS, \citealt{HEALEY2008}) with the 4FGL and found that only 858/1625 candidates were detected by \emph{Fermi}/LAT. Then, they selected eleven of these non-gamma-ray blazars and performed SED modeling with a one-zone leptonic model. They suggested that the detection of gamma rays might depend on the region where most of the jet power is dissipated, which is quite far from the central black hole for non-gamma-ray blazars and therefore unlikely to be compact enough to allow the escape of high-energy gamma rays \citep{DENG2021}. However, this condition can change, the dissipation region can drift outward or inward, as we have already observed in some cases (e.g. \citealt{FOSCHINI2011,GHISELLINI2013}).

\begin{figure}[!t]
\begin{center}
\includegraphics[scale=0.3]{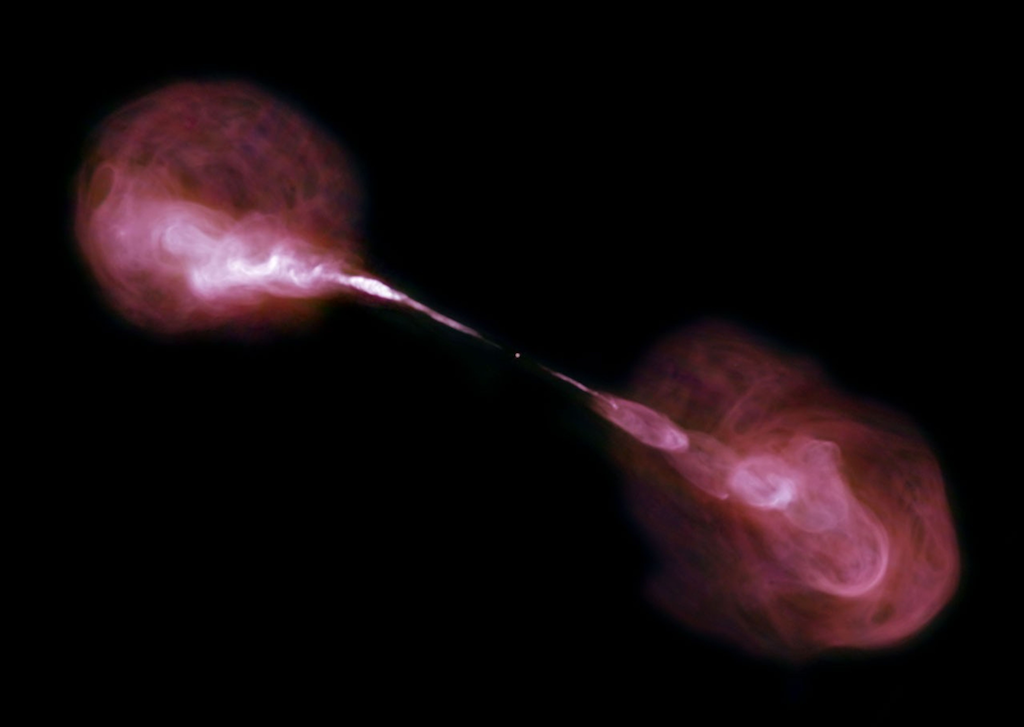}
\caption{Very Large Array (VLA) observation of the radio galaxy 3C~348 (a.k.a. Her A), done from August 2010 through September 2011, at $4-9$~GHz. From \href{https://science.nasa.gov/asset/hubble/vla-radio-image-of-hercules-a-3c-348/}{NASA Press Release}.}
\label{fig:herculesA}
\end{center}
\end{figure}

Yet another hypothesis worth citing relies on the activity of the central engine: \cite{CAPETTI2011} reported three cases of jetted AGN in which the central engine shut down, leaving the jets as radio relics (3C~28, 3C~314.1, 3C~348). The smoking gun could be the lack of [OIII] lines in the optical spectra, but not the H$\beta$ narrow component, since the oxygen recombination time is faster than that of the hydrogen. It is obvious that these latter objects will never be detected at gamma rays despite the presence of enormous radio jets (Fig.~\ref{fig:herculesA}), at least until the central engine is re-ignited.   

\section{Final Remarks}
Nine years ago, I wrote an essay titled ``What we talk about when we talk about blazars?'' \citep{FOSCHINI2017}. Today, I could write: what we talk about when we talk about Seyferts? It is very instructive to read papers of decades ago to understand that things that are taken for granted today, were not at all so back then, and to understand the semantic shift of many astrophysical terms. Indeed, even though the meaning of certain terms has changed, people's vocabulary hasn't. And this is one of the problems, because the permanence of a word, even in scientific jargon, doesn't guarantee the continuity of the concept. On the other hand, memory impairment can lead to ingenious constructions. Even those uninterested in classification cannot help but adopt a basic understanding of the terminology, even if they don't delve into the nuances of individual classes or subclasses. This can therefore generate interesting study paths based on a skill gap, even if one wishes to remain within the standards. 

The main obstacle to a classification of AGN is the difficulty to find a term that takes into account the temporal changes of the object to which it refers, because AGN are not static and immutable objects. They change, they evolve with time, moving from one class to another. There is the changing-look phenomenon, but also the cosmological evolution. A quasar is different from a BL Lac object, but it is still the same AGN that has evolved over time, just like I was different when I was 20 compared to today, but it is still me!

It is still relevant what Lodewijk Woltjer wrote in 1990: ``active galactic nuclei constitute a somewhat vaguely defined class of objects'' \citep{BLANDFORD1990}. Despite the unification model of AGN \citep{ANTONUCCI1993,URRY1995}, still today we use classifications depending on appearance and the adopted instruments, which is the worst way to classify any object, as noted already by \cite{GASKELL1987}. Therefore, an object observed at radio frequencies can have flat or steep spectrum, jets or not, with a wide variety of shapes. The same object observed by using an optical spectrograph is classified according to the line shapes as NLS1, Seyfert 1, intermediate, Seyfert 2. If it is far enough, it becomes a quasar or a quasi-stellar object (QSO). Consequently, the classification will reflect the quantities and structures observed at those frequencies/wavelengths and can \emph{never} provide a complete, though not exhaustive, description of the physical nature of the object.

Furthermore, advances in technology can make certain definitions based on instrumental capabilities obsolete. Therefore, if 30-40 years ago it was almost impossible to observe the host galaxy at $z\gtrsim 0.1$, today, with the \emph{James Webb Space Telescope} we are exploring the far Universe. If 30-40 years ago 5\AA\, of equivalent width (EW) could be a sufficient threshold to have a robust detection of an emission line, today we can go down to m\AA. The availability of large surveys has opened the door of the time-domain astronomy, and we have come to understand that objects belonging to different classes are actually the same, but observed at different times, not only from a different viewing angle. Therefore, the same AGN can display emission lines with EWs spanning on a large interval of values, depending on the activity of jet or the disk accretion rate. Although the unified model is geometric and static, today we can update it by adding the temporal coordinate, as Woltjer already warned more than thirty years ago \citep{BLANDFORD1990}.

\begin{figure}[!t]
\begin{center}
\includegraphics[scale=0.15]{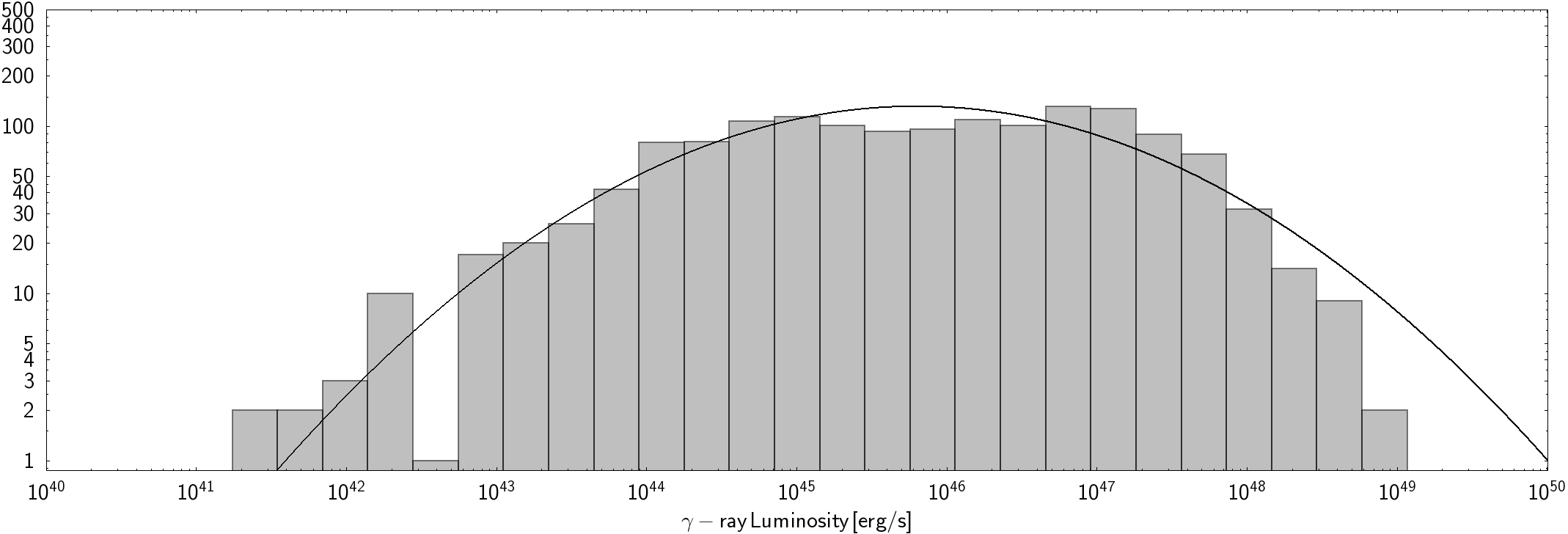}
\caption{This distribution of gamma-ray luminosity is the same of Fig.~\ref{fig:gammapow}, but without the different classes. In this case, I kept also CLAGN and AMB classes, which are not present in Fig.~\ref{fig:gammapow}.}
\label{fig:lumgammadistr}
\end{center}
\end{figure}

Despite all its limitations and flaws, the unified model holds up well and tells us that the central engine is always the same, so one might ask: why continue to use so many labels for the same object? Fig.~\ref{fig:lumgammadistr} displays the distribution of the gamma-ray luminosity of the 1477 jetted AGN of the rev4FGL with a spectroscopic redshift. It is the same of Fig.~\ref{fig:gammapow}, but I simply removed the symbols of different classes. One could see just a gaussian distribution with a low-luminosity tail, and an intriguing gap between $\sim 10^{42-43}$~erg~s$^{-1}$. One would expect a continuous transition from blazar to radio galaxies, because of the change of the viewing angle, but this gap suggests that there might be something else, and calls for a deep study of the gamma-ray emission of low-luminosity AGN.

Let's put this tail and gap aside and ask: what is the difference between Fig.~\ref{fig:gammapow} and Fig.~\ref{fig:lumgammadistr}? We have lost information. We have lost the information about the discovery of gamma-ray emission from NLS1s, which instead included them among the most powerful jetted AGN, comparable to blazars. The relativistic jets previously discovered in radio observations (e.g. \citealt{ZHOU2003,DOI2006,DOI2007}) have therefore proven -- at least in some cases -- to be active and powerful enough to emit high-energy gamma rays. Therefore, they are not relics from remote epochs, or due to starburst activity, or a combination of accretion mode and spin (e.g. \citealt{KOMOSSA2006}). By using the luminosity function, \cite{BERTON2016} proved that NLS1s are the low-luminosity tail of FSRQs, and so one might question the need to keep the two classes separate. However, NLS1s have central black hole masses smaller than FSRQs, have comparable, if not higher, accretion rates, and are mostly hosted by disk galaxies. If we had continued to think of NLS1s as FSRQs, we would not have noticed these differences and would still believe today that to generate powerful relativistic jets, supermassive black holes with a mass greater than $\sim 10^{8}M_{\odot}$ hosted in elliptical galaxies are needed. We wouldn't have understood what was wrong with that paradigm. Or maybe we would have understood it, sooner or later, but in these conversations it is always good to remember what Galileo Galilei said: 

\begin{quotation}
``All truths are easy to understand once they are discovered; the point is to discover them''.
\end{quotation}

In Fig.~\ref{fig:lumgammadistr}, we have also lost information on the detection of gamma-ray emission of other Seyfert subclasses (type 1, intermediate, 2), which instead paved the way for the understanding that disk winds can also emit gamma rays \citep{LENAIN2010,MICHIYAMA2024,PERETTI2025}, which in turn offers interesting prospects for finding counterparts to neutrino detections \citep{PERETTI2025}.

Therefore, on one side (unified physical model), throwing everything into the same pot just ends up creating a muddy mess, thus missing various opportunities to improve our science. On the opposite side, despite all its limitations and flaws, classification based on observational appearances and instrumental performance responds to the primary linguistic need to give names to things in order to study them. It is therefore essential to avoid stopping at this initial, rudimentary classification based on appearances, believing it to be the key to understanding the nature of things. A physical classification is necessary, but can only be attempted after more in-depth study. The two types of classification are therefore complementary and not mutually exclusive.

Some years ago, I proposed a new more physical classification of jetted AGN to overcome the above cited problems of an observation-based classification \citep{FOSCHINI2017}, but today I understand that it is too limited, unsuitable, and almost useless. I think it would be better to update the scheme proposed by R. Blandford \citep{BLANDFORD1990}, linking the classification based on observation with that based on physical quantities, although some of these -- such as spin and viewing angle -- are still extremely difficult to measure today. Furthermore, that scheme lacks the temporal coordinate, which is perhaps impossible to insert. But we have to find a way to make this clear, even if I don't know how at the moment.

\begin{figure}[!t]
\begin{center}
\includegraphics[scale=0.15]{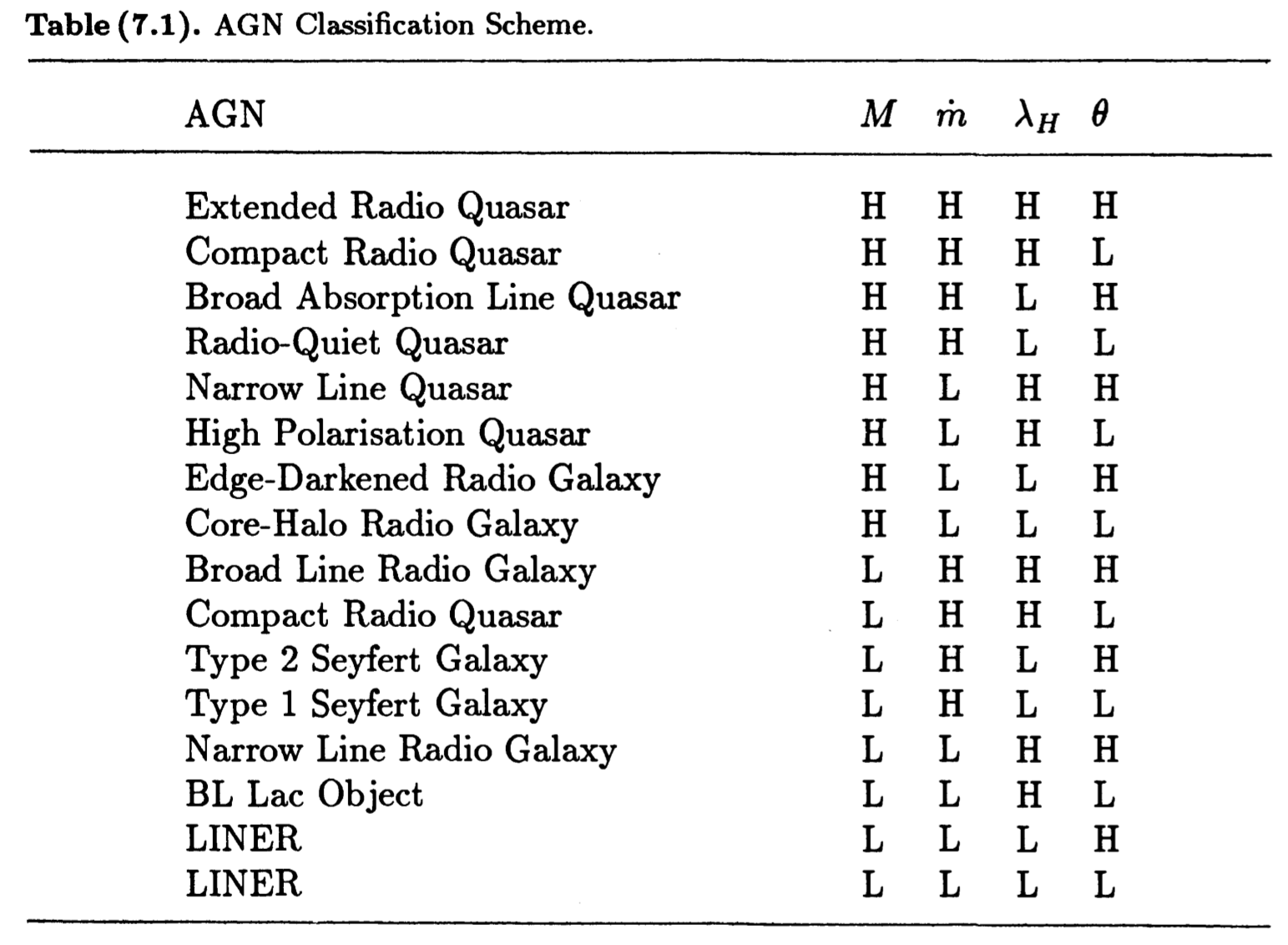}
\caption{Classification of AGN proposed by R. Blandford in \cite{BLANDFORD1990}. $M$ is the mass of the central black hole, $\dot{m}$ is the accretion rate, $\lambda_{\rm H}$ is the spin of the black hole, and $\theta$ is the viewing angle. The letters H and L mean high and low.}
\label{fig:agnblandford}
\end{center}
\end{figure}

\section*{Acknowledgements}
I would like to thank Guido Di Cocco for a useful and pleasant discussion on the MISO experiment and the early years of the gamma-ray astronomy.

This research has made use of data from the MOJAVE database that is maintained by the MOJAVE team \citep{LISTER2018}.

\end{document}